\documentclass{aastex6}

 \slugcomment{Accepted for publication in ApJ}
\shorttitle{Li-rich giants: SSC engulfment}
\shortauthors{Aguilera-G\'omez et al.}

\begin{document}


\title{On Lithium-Rich Red Giants. I. \\
    Engulfment of Sub-Stellar Companions}

\author{Claudia Aguilera-G\'omez\altaffilmark{1,2} and Julio Chanam\'e\altaffilmark{1}}
\affil{Instituto de Astrof\'isica, Pontificia Universidad Cat\'olica de Chile\\
 Av. Vicu\~na Mackenna 4860, 782-0436 Macul, Santiago, Chile}

\author{Marc H. Pinsonneault\altaffilmark{3}}
\affil{Department of Astronomy, The Ohio State University\\
 Columbus, OH 43210, USA.}
 

\and
\author{Joleen K. Carlberg\altaffilmark{4}}
\affil{NASA Goddard Space Flight Center\\
 Code 667, Greenbelt MD 20771 USA.}

\altaffiltext{1}{Millennium Institute of Astrophysics, Santiago, Chile}
\altaffiltext{2}{caguiler@astro.puc.cl}
\altaffiltext{3}{Center for Cosmology and Astroparticle Physics, The Ohio State University, Columbus, OH 43210, USA.}
\altaffiltext{4}{NASA Postdoctoral Program Fellow.}

\begin{abstract}

A small fraction of red giants are known to be lithium (Li) rich, in contradiction with expectations from stellar evolutionary theory. A possible explanation for these atypical giants is the engulfment of a Li-rich planet or brown dwarf by the star. In this work, we model the evolution of Li abundance in canonical red giants including the accretion of a sub-stellar mass companion. We consider a wide range of stellar and companion masses, Li abundances, stellar metallicities, and planetary orbital periods. Based on our calculations, companions with masses lower than $15\,\mathrm{M_J}$ dissolve in the convective envelope and can induce Li enrichment in regimes where extra mixing does not operate. Our models indicate that the accretion of a substellar companion can explain abundances up to A(Li)$\approx$2.2, setting an upper limit for Li-rich giants formed by this mechanism. Giants with higher abundances need another mechanism to be explained. For reasonable planetary distributions, we predict the Li abundance distribution of low-mass giants undergoing planet engulfment, finding that between 1$\%$ to 3$\%$ of them should have $\mathrm{A(Li)}\geq1.5$. We show that depending on the stellar mass range, this traditional definition of Li-rich giants is misleading, as isolated massive stars would be considered anomalous while giants engulfing a companion would be set aside, flagged as normal. We explore the detectability of companion engulfment, finding that planets with masses higher than $\sim 7\,\mathrm{M_J}$ produce a distinct signature, and that descendants of stars originating in the Li-dip and low luminosity red giants are ideal tests of this channel.

\end{abstract}


\section{Introduction}

Lithium (Li) is a very fragile element and one of the few that is synthesized almost immediately after the Big Bang \citep{wag67}. Li is also used extensively as a proxy for mixing inside stars, 
because its abundance is very sensitive to the environmental conditions \citep[e.g.,][]{ws65, mich91, chabo95, marc97, ses05}. It is produced inside stars by 
nuclear burning and is rapidly destroyed by proton capture at temperatures 
($T\gtrsim2.5\times10^6$) and densities that are reached in the inner regions of low mass stars during the main sequence (MS). 
The abundance of Li can also change depending on processes acting inside the star that are not typically included in standard stellar models, such as rotation, microscopic diffusion, mass loss and mass accretion.

Canonical stellar evolution predicts solar-type MS stars with Li depleted interiors, but no surface depletion during the MS and no dispersion at fixed mass, metallicity and age. Yet, 
there is a large spread in Li abundances for stars in the MS with similar atmospheric 
parameters \citep[e.g.][]{lr04}, indicating that even for solar type stars some non-standard extra 
mixing mechanism is acting, like rotation \citep[e.g.][]{pin92}, mass loss \citep[e.g.][]{sf92}, diffusion \citep[e.g.][]{mich86} or gravity waves \citep[e.g.][]{ct05}. 
This means that the surface abundance of MS stars does not reflect the initial Li abundance. In the Sun, for example, the current atmospheric abundance \citep[$\mathrm{A(Li)}_{\odot}$\footnote{$A(Li)=log(N_{Li}/N_{H})+12.00$, where $N_x$ is the number of atoms of element ``$x$"}$=1.05$,][]{asp09} is more than 2 dex lower than the meteoritic abundance 
($\mathrm{A(Li)}=3.3$), and well below solar analogs in young star clusters.

The distribution of Li inside the stars changes as 
the stars ascend the red giant branch (RGB). The Li remaining on the outer layers of the star is diluted  when the surface convection zone deepens, mixing the nuclear 
processed material from the interior with material from the surface of the star, in the process known as first 
dredge up (FDU). Another episode of surface abundance changes occurs for low mass stars at the moment of the red giant branch bump, the phase where the hydrogen burning shell crosses the discontinuity in the mean molecular weight left by the deepening convective envelope. Prior to this, some authors \citep[e.g.,][]{den03} suggest that strong $\mu$ gradients below the convection zone and between it and the hydrogen burning shell will strongly suppress the mixing. According to \citet{julio05}, though, rotational mixing in the lower RGB phase is inefficient independent of whether there is a $\mu$-barrier. How this episode of extra mixing occurs and its overall effects on the surface abundances are still debated in the literature, but there is strong observational evidence that the removal of the chemical discontinuity would allow extra-mixing that carries Li depleted material into the stellar surface, further decreasing the Li abundance \citep{lind09,pal11} and also producing alterations on the surface abundances of C, N, O and the carbon isotopic ratio \citep[e.g.][]{spi05,gra00, she03}. This non-convective, universal mixing in low mass stars after the RGB luminosity function bump is what is called ``canonical extra-mixing" \citep{den03}, and it seems to be much more efficient for low metallicity stars.

There has been a considerable body of work on possible mixing mechanisms inside stars. One of those is rotationally induced 
mixing \citep{dh04}, but the results regarding this process are not conclusive. \citet{julio05} used a maximal mixing approach and showed that rotationally induced mixing cannot reproduce the spectrum of observational evidence using reasonable rotation rates. Another potential mechanism
is thermohaline mixing \citep{egg06,cz07} where the reaction ${}^3$He(${}^3$He, 
2p)${}^4$He creates an inversion of molecular weight gradient with depth, triggering an instability analogous to salt-finger ones in the ocean. Again, how this type of mixing affects the CN abundance of giants is not known \citep{can10, ang12}. Among other extra-mixing 
mechanisms, we can also name magnetic buoyancy (\citealt{bus07, nord08}) and Magneto-thermohaline mixing (\citealt{den09}).

The Li abundance of a star that begins its RGB phase with the solar meteoritic abundance should therefore be $\sim1.8$ dex lower than the MS turn-off value, setting an upper limit of $\mathrm{A(Li)}=1.5$ for what the 
Li abundance is expected to be in these giant stars. In general, we expect much lower Li in the RGB phase, with actual dilution factors from $\sim30$ to $\sim40000$, depending on the stellar mass, metallicity and also considering Li burning during the FDU. Although this expected low surface Li abundance is what is observed for most of the G-K giants, agreeing 
with the canonical prescription for stellar evolution, an interesting percentage of giants have $\mathrm{A(Li})>1.5$, defining what is usually considered a Li enriched giant star. These unusual giants were  discovered by \citet{ws82}, but the number of detections has 
greatly increased since then, in 
environments ranging from clusters to the field \citep[e.g.,][]{das95, bala00, rl05, gon09, ruch11,mon11,leb12,kir12,ms13, dm16}. Li-rich giants are now recognized as a real, but uncommon population representing about 1-2\% of the giants, depending on the sample \citep[e.g.][]{bro89, kum11} and location. They seem to be extremely rare in globular clusters \citep[][]{Kir16} where only a few Li-rich giants are known \citep[e.g.][]{dor15}.

Several explanations have been proposed for these puzzling giants. Some of them, which we refer to as 
internal mechanisms, require the production of Li inside the star via the Cameron-Fowler nuclear chain \citep{CF71}, where the star's $^3$He is converted into $^7$Li by first producing beryllium 
($^3$He($^4$He,$\gamma$)$^7$Be), which then decays into Li by electron capture ($^7$Be($e^-$,$\nu$)$^7$Li). These reactions 
are a branch of the well known p-p chain of hydrogen fusion, in which the $^7$Li usually gets 
rapidly destroyed by capturing protons to finally produce $^4$He nuclei. Therefore, in order to survive once created, the freshly 
generated Li or beryllium must be quickly brought to a zone of the star with cooler temperatures, where 
Li is not immediately burned. 
This occurs naturally in the AGB phase of some stars during shell flashes \citep[via hot bottom burning,][]{sb92}, 
where the beryllium can be produced under convective conditions and therefore gets transported to cooler regions, so that when it decays into Li the element is preserved and does not burn by proton 
capture.
In the case of the Cameron-Fowler mechanism acting during the RGB phase of low mass stars \citep{sb99}, an extra mixing process is needed to connect the convective envelope to the H-burning zone, 
allowing the material to circulate and deposit $^7$Be in cooler regions, where it can produce a visible 
signature in Li abundance. Then, the internal scenarios differ only in the way of transporting the Li or 
parent Be into the base of the convective layer.

As with the canonical extra-mixing, several possibilities have been suggested for this ``enhanced extra-mixing" that could contribute to explain Li-rich giants. Related to rotationally induced mixing, \citet{dh04} relied on the erasing of the discontinuity in mean molecular weight to produce short lived increases in Li abundance, but, consistent with the results from \citet{julio05}, \citet{pala06} find that meridional circulation and shear instabilities do 
not get high enough mixing rates to reproduce the anomalous observed abundances and to increase the amount of Li in the envelope of stars. Thermohaline mixing was another possibility, but it seems that the aspect ratio of the fingers producing the mixing is not high enough to explain observations \citep[][see \citealt{gar15} for a discussion]{den10,wau14}, and to explain Li-rich giants the mixing efficiency would need to be extremely and unrealistically high \citep{den10}. This is why other possibilities have been explored, as magnetic buoyancy, which according to \citet{gua09}, should be able to produce Li enrichment up to $\mathrm{A(Li)}\sim2.5$.

 It is worth noticing that all of these extra-mixing mechanisms are expected to develop (or be much more efficient) after the RGB bump, thus there is a domain where the creation of Li-rich giants via internal Li-production plus mixing is not expected at all, and that is precisely before the bump and at higher metallicites where the observational evidence for extra-mixing is weaker.

The other kind of mechanisms proposed to explain Li-rich giants are those where the star obtains material 
from an external source. Objects with masses larger than $M\sim0.065\,\mathrm{M_{\odot}}$ will completely deplete the Li in their interiors \citep{cha96}, thus, a planet or a brown dwarf (BD) are still rich in Li and their engulfment could increase the superficial abundance of the star, as first 
suggested by \citet{ale67}. When stars evolve and expand their radii, it is natural to expect they will 
eventually engulf some of the planets they host. Which planets orbiting a certain star are in jeopardy 
is a matter of debate that not only involves how much the radius of the host star grows, but also how the 
evolution of the orbit is treated and how tidal evolution is understood and modeled \citep[][among others]{vill09,kuni11,vill14}. And there is also additional complications due to where do the planets dissolve. Regardless of which planets in each system end up being consumed by their host star, this is a process known to occur and in some special cases there is even evidence pointing to a recent ingestion. An 
interesting case is that of the star BD+48 740, explored by \citet{ada12}.

Since extra-mixing develops in the giants after the luminosity function bump, if we were to look for Li produced exclusively by substellar companion ingestion, without the added complications of dealing with mixing, we would have to look for a signal that occurs after the FDU and before the RGB bump. Added to this, to see only the signal due to planet or brown dwarf engulfment, it would be ideal to focus on stars that left the MS with very little Li, such as stars in the Li-dip \citep{bo86, bala95,dm15}. Even if the signature produced in these is smaller, the lack of contamination due to other giants would allow to isolate the effects of companion accretion. The substellar companion engulfment scenario is also an interesting potential test of planetary populations in mass regimes where characterization by radial velocity is difficult.

Planets and brown dwarfs are not the only possible external source of Li. If the RGB star happens to be the binary companion to an AGB star that has already undergone a process of Li-enrichment, then this more evolved star could transfer part of its Li-enriched material to the red giant \citep{sb99}. Even if this process is an interesting alternative, only the higher mass AGB are affected by the Li-production process and these are also found in a very specific metallicity domain.

While rare in terms of their population, Li rich giants challenge the canonical theory of stellar 
evolution, and a satisfactory understanding of this anomaly could help us comprehend non-standard 
ingredients involved in the evolution of stars, as planet-star interaction, extra mixing, and light element 
production.

In the present paper, with the goal of constraining the mechanisms that can produce Li rich giants, we 
model the engulfment of a sub-stellar companion (SSC) by its host star using reasonable assumptions. We explore the available parameter space of both stellar and companion properties. Our goal 
is to determine whether SSC engulfment is enough to explain all Li rich RGB stars, or only a subset of the 
observations, 
and which.

The accretion of SSCs has been modeled by \citeauthor{sl99a} \citep{sl99a,sl99b}, who calculated 
the observational signatures of engulfment by an AGB star of $2.9\,\mathrm{M_\odot}$ \citep{sl99a} and RGB 
stars of masses close to $1.0\,\mathrm{M_\odot}$ \citep{sl99b}, including Li enhancement, fast rotation and 
infrared excess. These models are limited, as they do not account for the most typical masses
of RGB stars. Here, 
using more current tools we overcome some of these limitations, modeling the process of SSC engulfment for a 
larger range of parameter space, only focusing on Li enhancement, and testing 
different initial conditions that would produce the most optimistic case scenario. This also allows us to predict the population of Li-rich giants that we could find in domains where other type of mechanisms currently explored are not efficient in producing Li.

The paper is organized as follows. Section 2 describes the models used, both for the star and SSC, and describes the assumptions implemented, also showing the point of SSC dissipation inside the stars. Our results regarding the increase in superficial Li abundance are presented in section 3, along with the rates of Li-enrichment due to SSC engulfment. Those results are discussed in section 4, to finally summarize our research in section 5.

\section{Models}

In order to study the process of SSC engulfment we need models for both the star and the accreted sub-stellar 
object. A self-consistent model of 
planet/BD engulfment would require a detailed study of the uncertain common envelope problem. Instead, we 
use a post-processing approach, where a grid of stellar evolutionary models is used as background points on 
top of which
the effects of SSC engulfment are added in a separate code. The advantage of this method is 
that 
the parameter space can be explored faster, and thus it is very useful when several cases need to be studied 
(as in this particular problem, where the parameters involved are stellar mass, stellar metallicity, SSC 
mass, 
SSC composition and the orbital period of the accreted object). The main disadvantage of this method is that 
we do not include feedback of the engulfment process on stellar evolution.
However, this is a reasonable starting point for our restricted problem, as we know that stars with a large range in masses have a similar dredge-up, so adding a small amount of Li to any star would represent most likely a perturbation to the FDU. Also, these models show a realistic case of where Li might be found as a starting point for more complicated later models. 

In this section we describe the grid of stellar evolutionary models used, as well as the characteristics and modelling of the SSCs.

\subsection{Star}\label{starsec}

We use the Yale Rotating Evolutionary Code \citep[YREC,][]{pin89, pin90,pin92,dem08}
to construct a grid of stellar evolutionary models, which do not include diffusion, 
rotation, overshooting or any other non-canonical ingredients. We use the mixing length theory for convection \citep{cox68}. Our models make use of the 2006 OPAL equation of  state \citep{rn02} and include 
nuclear reaction rates from \citet{adel11}, treating weak screening with the method of \citet{sal54}. Cross section for the $\mathrm{{}^7Li(p,\alpha)\alpha}$ is obtained from \citet{lam12}.
Also, high-temperature opacities from the Opacity Project are used \citep{men07}, complemented with \citet{ferg05} low temperature 
opacities. Atmosphere and boundary conditions are those from \citet{kur79}. See \citet{van12} for a summary of the input physics.

We study stellar models with masses between $M=1.0\,\mathrm{M_{\odot}}$ and $M=2.0\,\mathrm{M_{\odot}}$ in increments of $0.25\,\mathrm{M_{\odot}}$, and all of them are evolved up to the tip of the red giant branch. We adopt a current solar metal abundance $Z/X=0.02292$ from \citet{gs98} and choose the hydrogen mass fraction $X$ and mixing length coefficient $\alpha$ that reproduces the solar radius and luminosity for a $1\,\mathrm{M_{\odot}}$ model at an age of $4.57$ Gyrs, considering gravitational settling. The final calibrated values are $X=0.70751$, $\alpha=1.94661$ and $Z=0.01879$.
The chemical mixture for an arbitrary metallicity [Fe/H] is obtained by considering a linear relation between helium mass fraction with metal content
\begin{equation}
\label{eq1}
Y=Y_p+\frac{\Delta Y}{\Delta Z}Z,
\end{equation}
with $Y_{p}=0.2484$ the primordial helium mass fraction \citep{cy03}, and $\frac{\Delta Y}{\Delta Z}=1.3465$ the slope of the relation obtained by using both the solar calibrated composition and the primordial mixture in Equation (\ref{eq1}).
Based on the metallicity of most of the observed Li rich giants, the metallicity explored is in the range [Fe/H]$=-2.0$ and [Fe/H]$=0.0$, focusing specially on stars with metallicities closer to solar. This is equivalent to consider stars with mixtures $Z=0.00020$ and $Y=0.2487$ to $Z=0.01879$ and $Y=0.2737$ respectively.

The pre-MS depletion in low mass canonical models is less than the observed depletion in turn-off stars. We therefore reduce it to match the maximum value of Li turn-off abundance observed among clusters \citep[See][and references therein]{dob14}. We expect 3 different mass domains for the giants, each having a corresponding initial Li abundance, based on the trends of $\mathrm{A(Li)}$ with stellar mass at the turn-off of clusters at solar metallicity: Stars with masses $M>1.8\,\mathrm{M_\odot}$ could have a higher Li abundance, similar to the meteoritic value of $\mathrm{A(Li)}_{*,ini}=3.3$, while low mass stars, with $M<1.3\,\mathrm{M_\odot}$ correspond to an older population that would have a Li abundance at the turn-off from $\mathrm{A(Li)}_{*,ini}=2.3$ in the halo to $\mathrm{A(Li)}_{*,ini}=2.6$ at solar metallicity. There is also a range of masses where the abundances are mainly upper limits $\mathrm{A(Li)}_{*,ini}<1.0$, corresponding to stars that leave the MS without Li (Li-dip stars).

As a test case, we produce models with an initial Li abundance of $\mathrm{A(Li)}_{*,ini}=2.6$ for stars of all masses, to compare the signature of SSC engulfment without the effect of the different initial amount of Li. We test in Section \ref{i_lihost} how the turn-off Li changes the abundance post-accretion.

\subsection{Sub-stellar companion} \label{ssc_as}

Although YREC is currently capable of modelling low-mass objects like the SSCs we 
will consider, the use of such rigorous calculation is not necessary for this work. The engulfment process modeled in this work is designed to explote the upper envelope of the possible Li signature. Thus, the only parameters we will take into account are the mass and radius of the planet or BD and 
its composition, including the amount of Li.

The mass, radius and composition of planets are obtained from the grid of \citet{fort07}, for both giant planets and rock-ice-iron planets. For brown dwarfs instead we use the evolutionary models of \citet{bar03} to obtain mass and 
radius, and assume a composition similar to that of Jupiter and Saturn (Hydrogen and Helium mainly). In general, the mass of the accreted 
object goes from $1\,M_\oplus$ up to $0.065\,M_{\odot}$, from planets to brown dwarfs. Higher mass brown 
dwarfs will burn Li \citep{cha96}, so those are not considered. 

The precise models used for the SSC do not affect our final results. We tested this by using a different grid \citep{zs13} for rock-iron-ice planets and comparing to the results obtained with \citet{fort07}.

\subsection{SSC Li content}

To compute the stellar Li abundance after engulfment, we need the total content of Li of the SSC, i.e., the amount of Li mass that is accreted by the star. To know this, we fix the ratio between Li mass fraction and metals to the Solar System meteoritic value, assuming that SSCs formed will have the same fixed $\mathrm{X_{Li}/Z}$ but may have a different metal content $Z_{SSC}$, and consequently, a different mass fraction of Li. Then, we will just need the mass of the SSC to obtain the accreted Li content. 

The meteoritic Li mass fraction was calculated by using an abundance of $\mathrm{A(Li)}=3.3$ \citep{lod98}.  Planet composition is still an uncertain thing, even for those of the solar system, and several possibilities may be studied. We consider 3 mass domains for the SSC companion, where we choose the metal content accordingly. For brown dwarfs, Li is usually used to distinguish sub-stellar objects from low mass stars \citep[the ``lithium test",][]{reb92}, so we expect to find a significant and observable amount of Li in a large fraction of objects, but in every other sense, they have a very similar composition to stars. We use the primordial solar mixture, with a value for the overall planet metallicity of $Z=Z_{\odot}$. It may be possible to have low mass brown dwarfs with a higher metal content if we consider that some could be formed by mass accretion. We use this composition for SSCs with masses $15\mathrm{M_J}<M_{SSC}\leq 60 \mathrm{M_J}$.

Based on what is found in the Solar System, it seems that gaseous giant planets tend to have 
a composition more similar to the parent star
than rocky planets, which are over abundant in heavy, high condensation temperature 
elements \citep[e.g.,][]{pal00}. Given this apparent trend, the engulfment of a Jupiter-like planet may not be able to increase the surface Li abundance as much as it is needed to explain the existence of Li-rich giants. On the other hand, rocky planets 
that may have a higher Li abundance than giant planets, also have less mass so their signatures may also be too small to 
detect. Regardless of this possible difference, we will use the same Li abundance for all the SSCs modeled and different metal content for rocky and giant planets. Then, the second mass domain will be that of giant planets for which we will consider a metal content of $Z_{SSC}=2.5\,\mathrm{Z_{\odot}}$, based on what is found in Jupiter \citep{net12}. As these are formed by core accretion they should be slightly more metal rich than the star they orbit. The mass range of this type of planets is from $0.01\,\mathrm{M_J}$ to $15\,\mathrm{M_J}$.

At last, for lower mass planets, we consider a composition enriched in metals, with $Z=1$. This category includes Earth-like planets and every SSC with $M_{SSC}<0.01\,\mathrm{M_J}$. As we see, the high amount of metals for rocky planets will imply a higher mass of Li engulfed relative to the mass of the companion.

\subsection{Assumptions}

To simplify calculations we model the ingestion of the SSC using some assumptions and briefly discuss the effect of these assumptions. First, we are not taking into account the tidal evolution of the SSC. We will just determine a point of 
SSC dissipation and model the engulfment of planets with different MS periods up to the tip of the RGB, without worrying how the SSC got to that point. This allows us to 
explore the effects of the engulfment process without the need to include orbital evolution, a topic 
still debated that is beyond the scope of this paper and that is not relevant for our purposes of Li-enrichment.

Second, as soon as it is engulfed by the host, the SSC will be completely dissolved in a single point inside the star, 
determined by the characteristics of each system (See Section \ref{Pssc}). Thus, we 
are not considering an accretion rate or planet/BD mass loss before the complete dissolution of the object. The actual 
portion of the object that is dissolved in the convective zone depends on the masses of the SSC and the envelope of the star, and on more detailed planetary structure \citep{Sand98}. The accreted object may suffer partial evaporation 
of material before reaching that point, and maybe even before contacting the star itself. If evaporation inside the star is gradual the final Li incorporated is the same than dissolving the whole object in a point of the convective envelope. On the other hand, any mass stripped from the SSC before it reaches the star is lost for the purpose of Li enrichment. 

Another assumption in our models is the instantaneous mixing of chemical species in the convective zones of 
the star, that immediately homogenizes the material in those regions. This occurs because the time step
between models, corresponding to nuclear timescales, is longer than the convective turnover timescale. One final simplification that our post-processing approach implicitly assumes is that the engulfment of the SSC does not affect the structure of the star.

\subsection{Point of SSC dissipation} \label{Pssc}

In order to know where to deposit the SSC Li, we need to explore the location in which the SSC dissolves inside the star. Some companions will dissolve in the radiative interior of the star, and thus they will never produce an observable signature. 
Following the calculations of \citet{sl99a}, we identify the location where the planet or brown dwarf is 
dissolved inside the star as the radial distance from the stellar center where the ambient temperature is comparable to the virial temperature of the SSC
\begin{equation}
\label{eqTvir}
 T_{vir}\approx \frac{G\mu_{SSC} m_H M_{SSC}}{kR_{SSC}},
\end{equation}
where the subindex \textit{SSC} indicates the properties of the SSC, being $M$ the mass, $\mu$ the mean molecular weight, and $R$ the radius. The constants correspond to $G$ the gravitational constant, $k$ the Boltzmann constant, and $m_H$ the hydrogen mass.

At such a location the thermal kinetic 
energy of the star is comparable to the gravitational binding energy of the external object. We assume that the SSC is completely dissipated there. This again goes in favor of conserving any Li deposited by the SSC, since dissipation in deeper regions of the star would risk some Li being destroyed by the higher ambient temperature.
It is possible that kinetic orbital energy may change the exact point of SSC dissipation. During the orbital evolution some of that energy may be converted to thermal energy, increasing the ambient temperature and thus moving the locus of dissipation closer to the stellar surface. In our maximal lithium abundance approach, this difference, is not relevant as it does not change the final abundance nor the modeled scenarios. Moreover, an order of magnitude calculation shows that the effect of kinetic orbital energy is larger for less massive planets and it is practicable negligible for the range of SSC masses modeled throughout this paper.

The virial temperature depends on the intrinsic traits of the external object, characterizing it. It is relevant to 
notice that different SSCs can have the same virial temperature and will be evaporated at 
the same location of a given star. The locus of planet/BD dissipation is established when comparing the virial temperature to the temperature 
profile of the star, thus the same SSC, characterized by its virial temperature, will dissolve at a different 
location for each star modeled.

Because stars are centrally concentrated objects, tidal effects become important closer to the stellar core \citep{soker87} and could completely disrupt the SSC . The 
evaporation condition given by the virial temperature always occurs at a larger radius than that of the disruption via 
tides, so that when this effect becomes important the SSC has already dissolved into the stellar material. Even 
if this was not the case, the dissolution at an inner radius implies that the SSC material is kept in the 
radiative interior of the star, hence never been able to reach the observable outer convective layer (i.e., in the absence of non-canonical processes such as extra mixing, which are not considered here).

\begin{table*}[!htb]
\begin{center}
\caption{Example SSCs with their respective virial temperatures. \label{Examples}}
\begin{tabular}{lllll}
\tableline\tableline
Type of SSC & Mass ($\mathrm{M_J}$) & Radius ($\mathrm{R_\oplus}$) & Composition & Virial Temperature (K)\\
\tableline
Planet &$1.42\times10^{-4}$ &$0.48$ &Rock/Iron  &$7.609\times10^3$ \\
Planet &\tablenotemark{a}$3.15\times10^{-3}$ &$1.00$ &67\% Rock/33\% Iron &$7.089\times10^5$ \\
Planet &$5.35\times10^{-2}$ &$7.26$  & Hydrogen/Helium &$3.543\times10^4$ \\
Brown dwarf &$20$&$10.45$& Hydrogen/Helium &$9.644\times10^6$ \\
\tableline
Brown dwarf & $60$&$8.69$ &Hydrogen/Helium &$3.479\times10^7$ \\
Planet &$1.46$ &$12.96$  &Hydrogen/Helium  &$6.067\times10^5$ \\
Planet &$1.00\times10^{-3}$ & $0.77$ & Rock & $4.423\times10^5$\\
Planet &$3.15\times10^{-3}$&$0.77$ & Iron & $5.486\times10^5$\\
\tableline
\end{tabular}
\tablenotetext{$a$}{Equivalent to $1.00\,\mathrm{M_{\oplus}}$.}
\end{center}
\end{table*}

Using the models for planets and brown dwarfs, we outline in Table \ref{Examples} some example objects, their corresponding properties and their corresponding virial temperature. 

\begin{figure*}[!htb]
\begin{center}
\includegraphics[width=\textwidth]{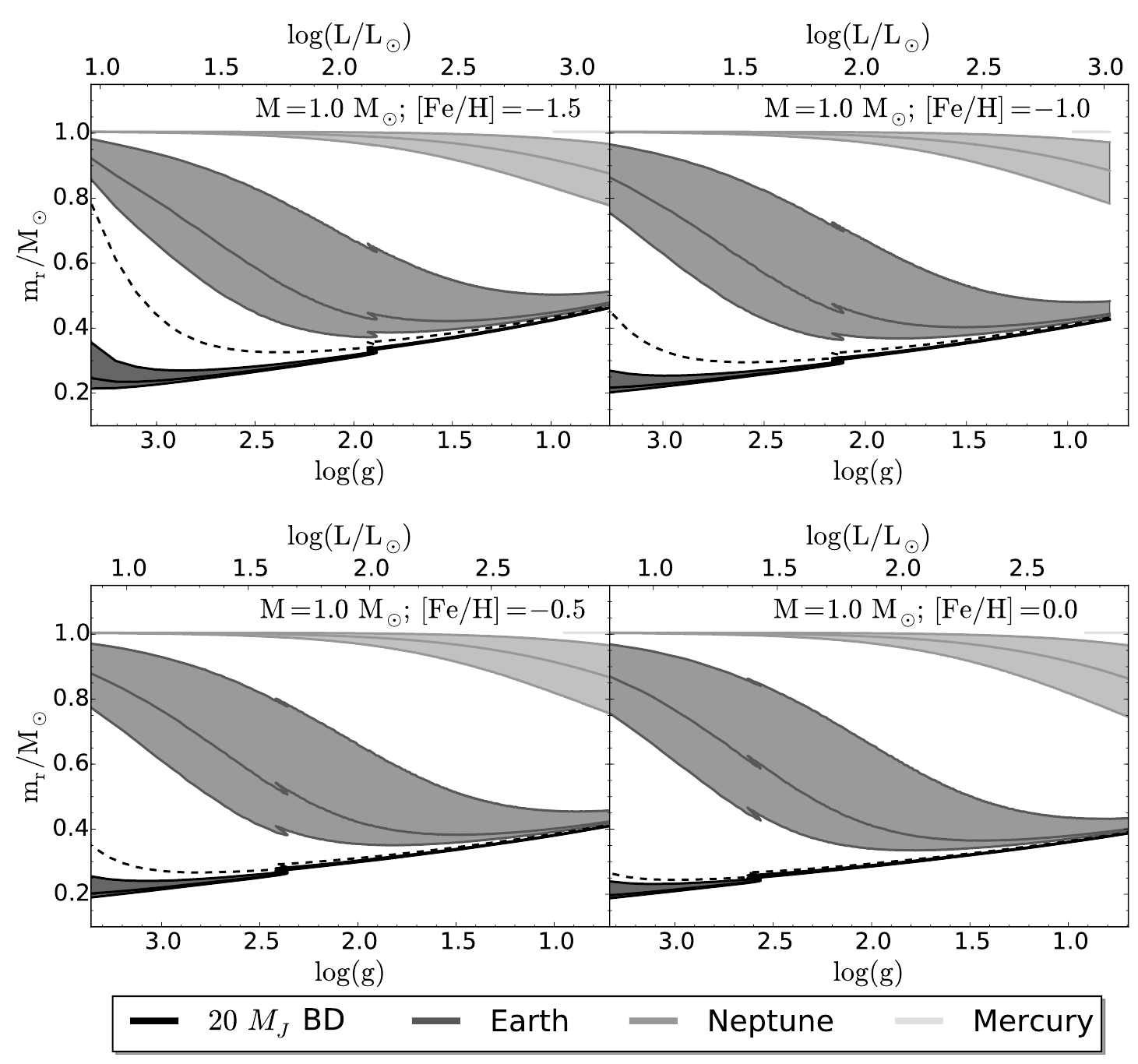}
\caption{Point of SSC dissipation for four different objects during the RGB 
phase of $1.0\,\mathrm{M_{\odot}}$ stars with different metallicities, as indicated on top of the panels. Each 
colored band represents the SSC whose virial temperature (with 5\% errors at either side) is indicated by the legend in the bottom. The black dashed line is the base of the convective envelope. In general, it can be seen that for different metallicities the point of SSC dissipation is very similar for a given luminosity, and that SSCs with larger virial temperatures will dissolve in the radiative region of the star.}
\label{Tvir}
\end{center}
\end{figure*}

The locus of SSC dissipation in $1.0\,\mathrm{M_{\odot}}$ stars with different metallicities can be seen in Figure \ref{Tvir} for a set of 4 different virial temperatures. 
These correspond to the first 4 virial temperatures shown in Table \ref{Examples} and were chosen because they cover the 
whole possible range of $T_{vir}$ values that can be obtained with very different SSC properties. The gray area 
around each line shows a $\sim 5\%$ value around each virial temperature in logarithm. The black dashed line is the base of the convective envelope of the star. It may be relevant to notice that a Jupiter-like planet would have a very similar $T_{vir}$ to that of Earth-like planets.

\begin{figure*}[!htb]
\begin{center}
\includegraphics[width=\textwidth]{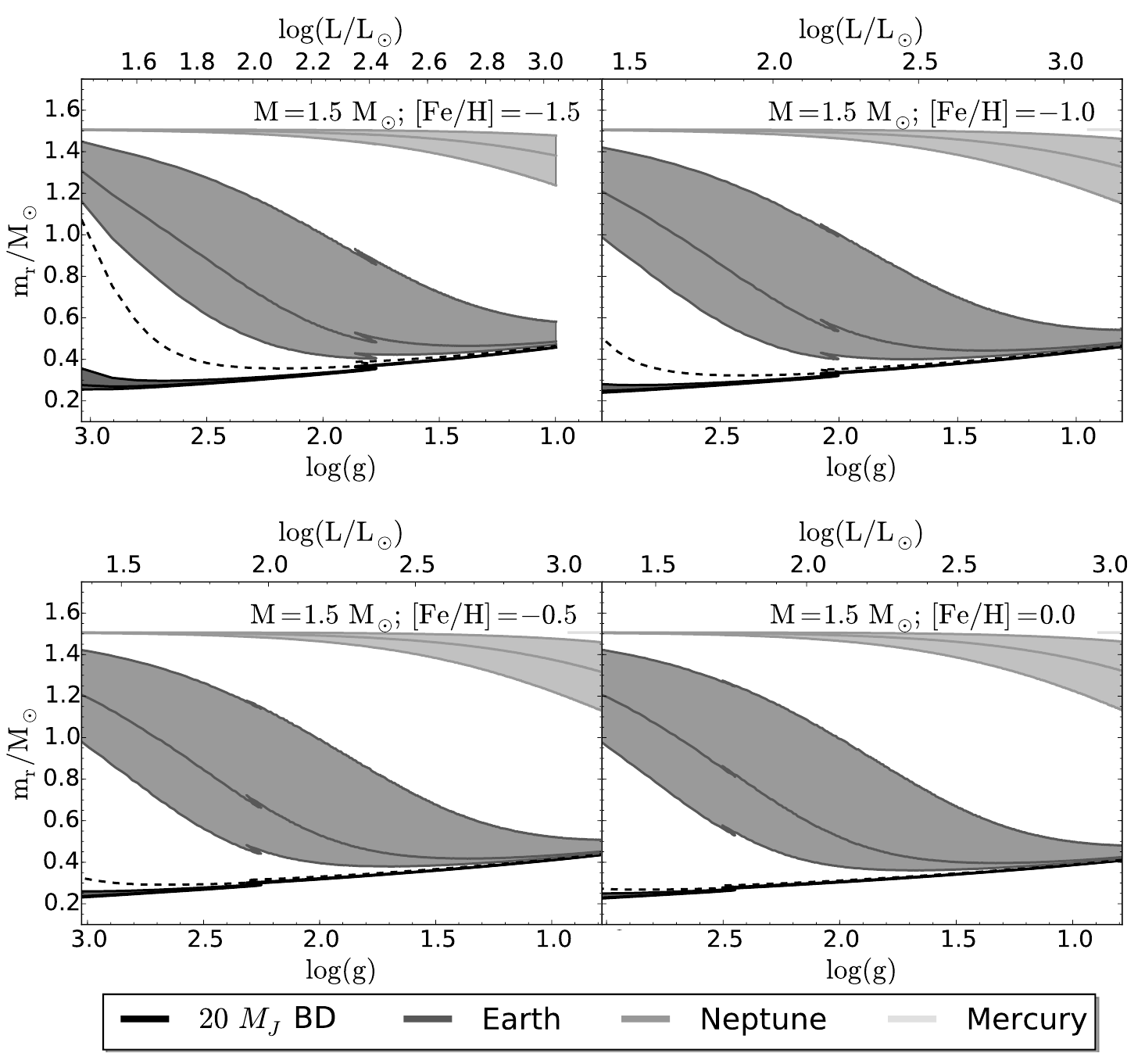}
\caption{Same as Figure \ref{Tvir} but for $1.5\,\mathrm{M_{\odot}}$ stars.}
\label{Tvir15}
\end{center}
\end{figure*}

The point of a given SSC dissipation at a certain luminosity does not move too much in mass coordinate for stars with the same mass but different metallicities, as can be observed when comparing the different panels in Figure \ref{Tvir}. Although SSC ingestion is our main goal and thus comparing point of dissipation is important, this figure also shows how the structural properties of the stars change with metallicity and allows to understand its effects on superficial Li abundance. The convective zone grows substantially more for metal rich stars, which we could expect since larger opacity makes radiative transport more difficult. This also occurs for stars of higher mass and in a similar metallicity range, as can be seen in Figure \ref{Tvir15}, which shows the point of SSC dissipation but for $1.5\,M_{\odot}$ stars with different metallicities. 

\begin{figure*}[!htb]
\begin{center}
\includegraphics[width=\textwidth]{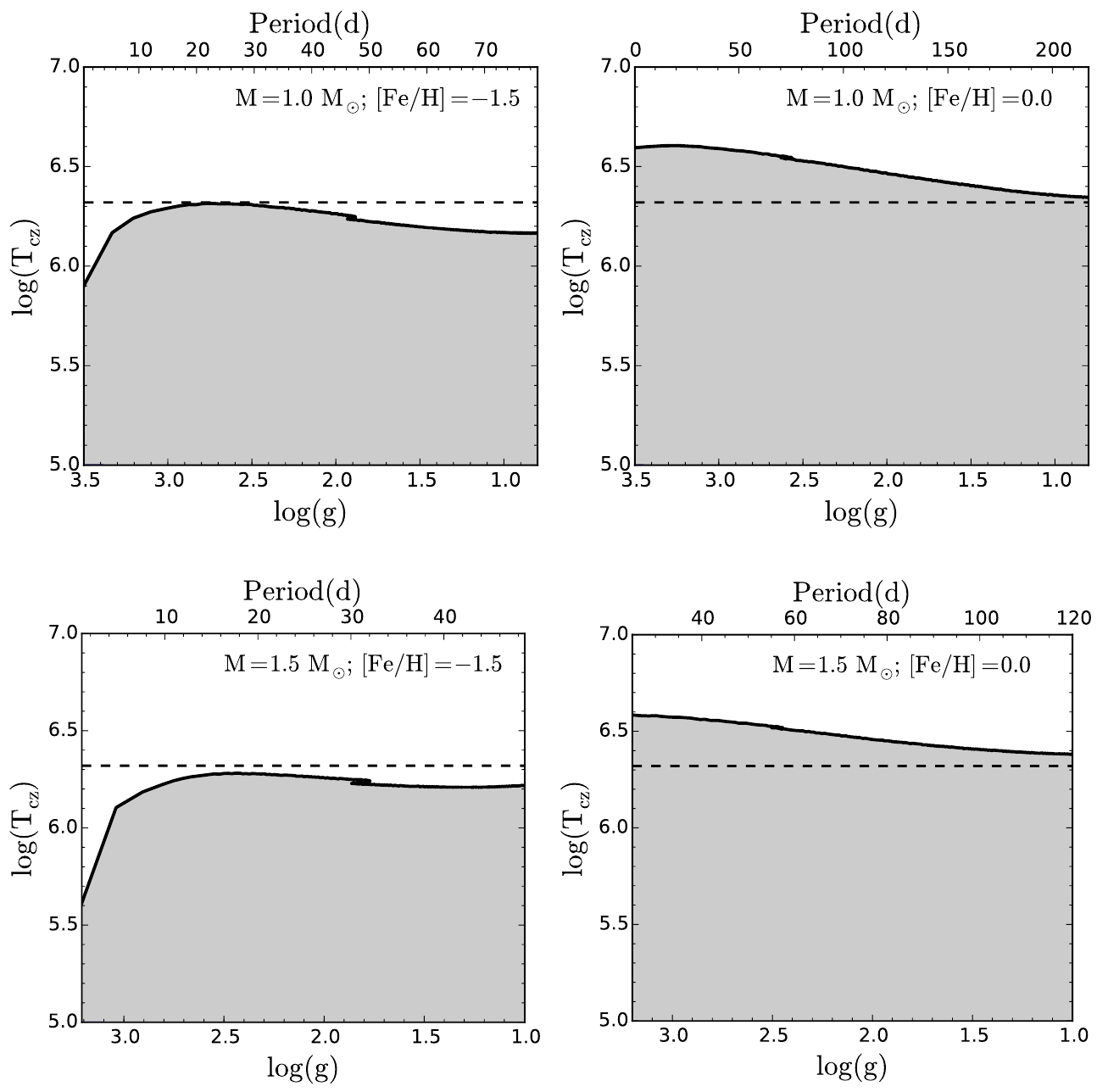}
\caption{Virial temperature at the base of the convective zone for stars of different masses and metallicites as a function of log(g) or period of engulfed planet. Here we can identify the SSCs that will end up dissolving in the convective envelope (shaded region) or in the radiative interior. The dashed horizontal line in each panel represents a low-mass brown dwarf of $15\,\mathrm{M_J}$.}
\label{Wheredis}
\end{center}
\end{figure*}

Even if the convective region reaches hotter temperatures for higher metallicity stars, this does not necessarily mean that the Li is going to burn under convective conditions. The condition usually given as a Li-burning limit is a temperature of $T=2.5\times10^6$ K, but this condition is not precise during the RGB phase of the stars, given the lower densities present. Consequently, a region of the star with higher temperatures may or may not burn the lithium based on its other properties.

Figure \ref{Wheredis} shows the limiting virial temperature between convective (shaded region) and radiative zone for stars of different masses and metallicities. Here it is clear that SSCs with a large virial temperature, which means companions with higher masses and/or smaller radii, will dissolve in the 
radiative region of the star, and as such, they cannot produce an observable 
signature in their surface Li content without extra mixing. Internal mechanisms of extra mixing are beyond the scope of the present paper and thus we will not study those objects in detail, and will only focus on SSC dissolved in the convective envelope.
As this Figure shows, given a fixed stellar mass, the radius of the star is a function of metallicity. This implies that SSC engulfed at the same $\log(g)$ in different panels are not the same.

A dashed horizontal line in each panel of Figure \ref{Wheredis} shows the virial temperature of a $15\,\mathrm{M_J}$ brown dwarf. It is clear that objects much more massive than this will dissolve always in the radiative interior of the star. Then, we can set an upper bound of $15$ to $20\,\mathrm{M_J}$ for the mass of the companion that would dissolve in the convective zone, always considering that the exact limiting mass of the SSC depends on the metallicity of the star. Even if one could have naively expected that the most massive SSCs, that have the largest Li content, would produce the largest signal in lithium abundance post-engulfment, this is not the case, as those very massive objects would not even produce a signature.

\section{Results: Evolution of the superficial Li abundance}
We will begin presenting our results for engulfed SSC of $\sim 15\,M_J$ and $Z_{SSC}=2.5\,\mathrm{Z_{\odot}}$, a realistic model for a gas giant planet with a high mass (or a brown dwarf with low mass). We will show results with different SSC properties and discuss differences in Sections \ref{i_lissc} and \ref{i_lihost}. For the star, the initial lithium abundance used is $\mathrm{A(Li)_{*,ini}}=2.6$.

\subsection{Engulfment of one SSC}

In this section we proceed to dilute the SSC inside the convective envelope and add its Li to that of the star, thus following the evolution of Li during the RGB phase. 

If we assume this dilution is instantaneous, so that the envelope mass ($M_{env}$) does not change due to standard RGB evolution (i.e., it does so only due to the incorporation of the SSC) and no nuclear burning is taking place in the convection zone, simple dilution gives the new mass fraction of any element $X_i$ \citep[equation 2,][]{sl99b}:

\begin{equation}
X_i=\frac{X^{env}_i \times M_{env}+X^{acc}_i \times M_{acc}}{M_{acc}+M_{env}},
\end{equation}
where $X_{env}$ and $X_{acc}$ are the mass fractions of the element in the envelope of the star and SSC respectively. In our models the accreted mass $M_{acc}$ is the whole mass of the SSC. Once the companion is diluted and its content added to that of the star, the model allows the burning of planetary Li, according to the burning rates inside the star.

\begin{figure*}[!htb]
\begin{center}
\includegraphics[width=\textwidth]{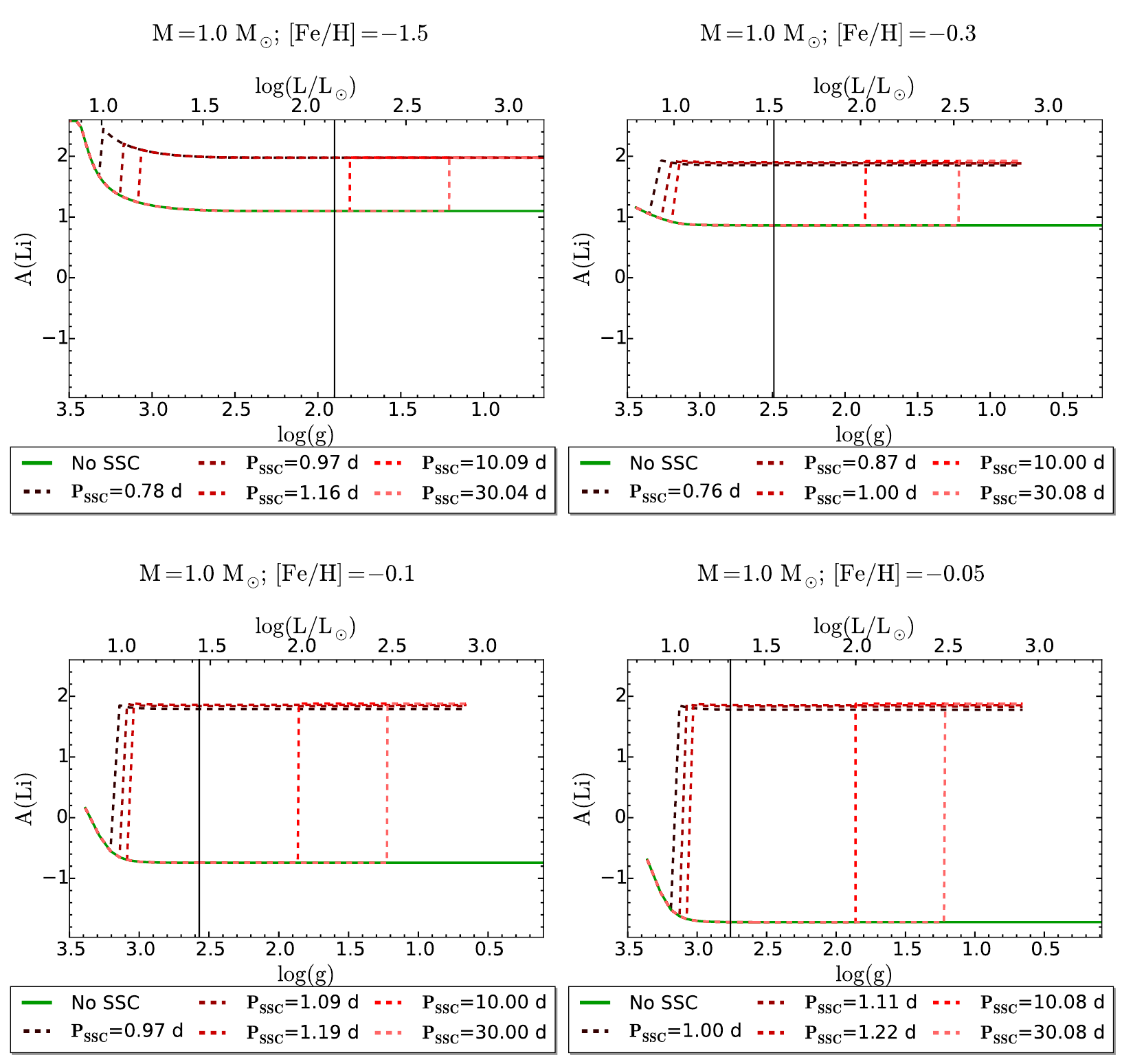}
\caption{Evolution of the superficial Li abundance for a $1.0\,\mathrm{M_{\odot}}$ with $\mathrm{A(Li)_{*,ini}}=2.6$ and different metallicities, as indicated for each panel. The green solid lines indicates A(Li) for the star evolving in isolation while the red dashed lines are the A(Li) after planet engulfment for planets with different periods $P_{SSC}$ in days. As a reference, the evolution is shown as a function of either log(g) or as a function of luminosity and the luminosity function bump is depicted in each case as a black solid vertical line.}
\label{Lisup}
\end{center}
\end{figure*}

\begin{figure*}[!htb]
\begin{center}
\includegraphics[width=\textwidth]{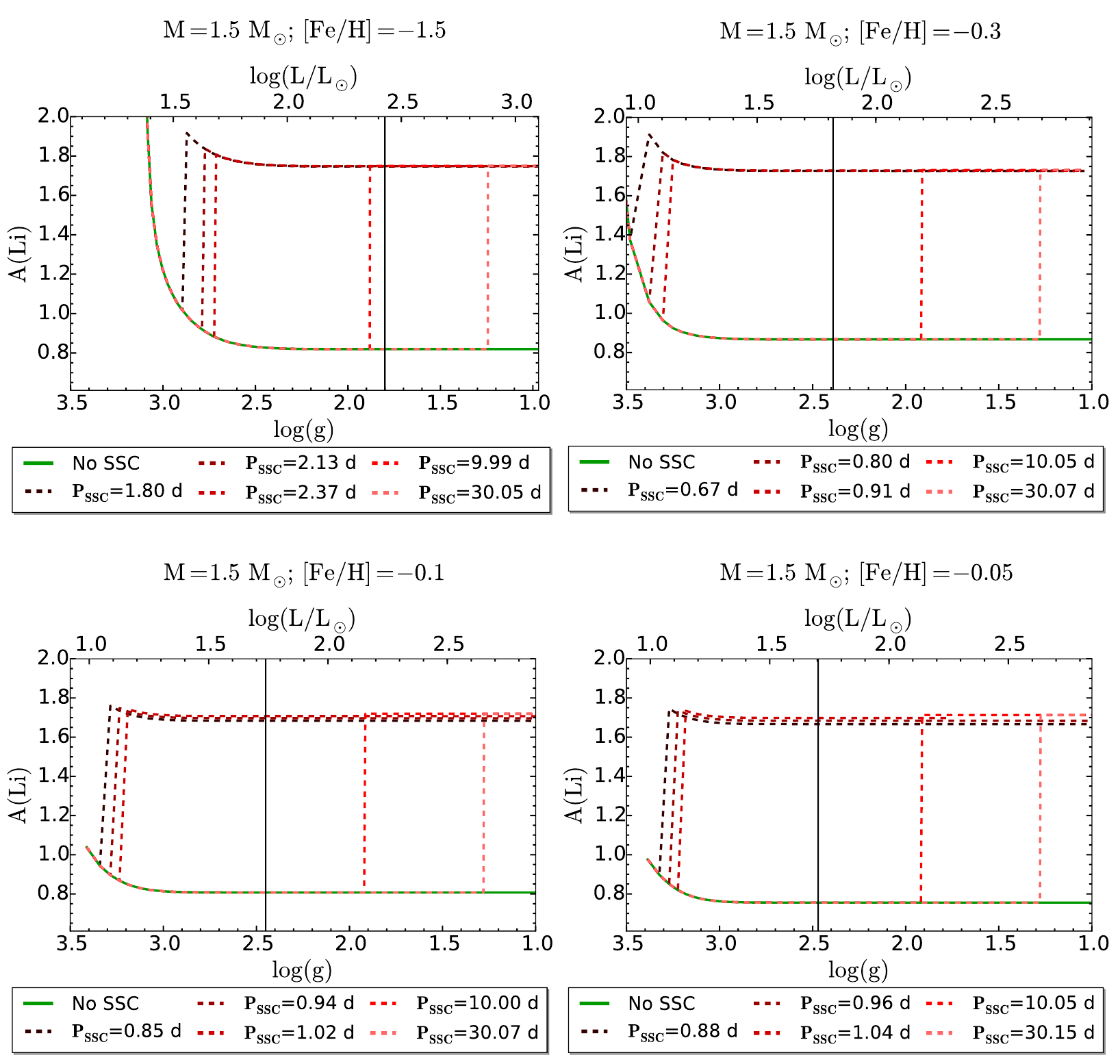}
\caption{Same as Figure \ref{Lisup} for a $1.5\,\mathrm{M_{\odot}}$.}
\label{Lisup15}
\end{center}
\end{figure*}

\begin{figure*}[!htb]
\begin{center}
\includegraphics[width=\textwidth]{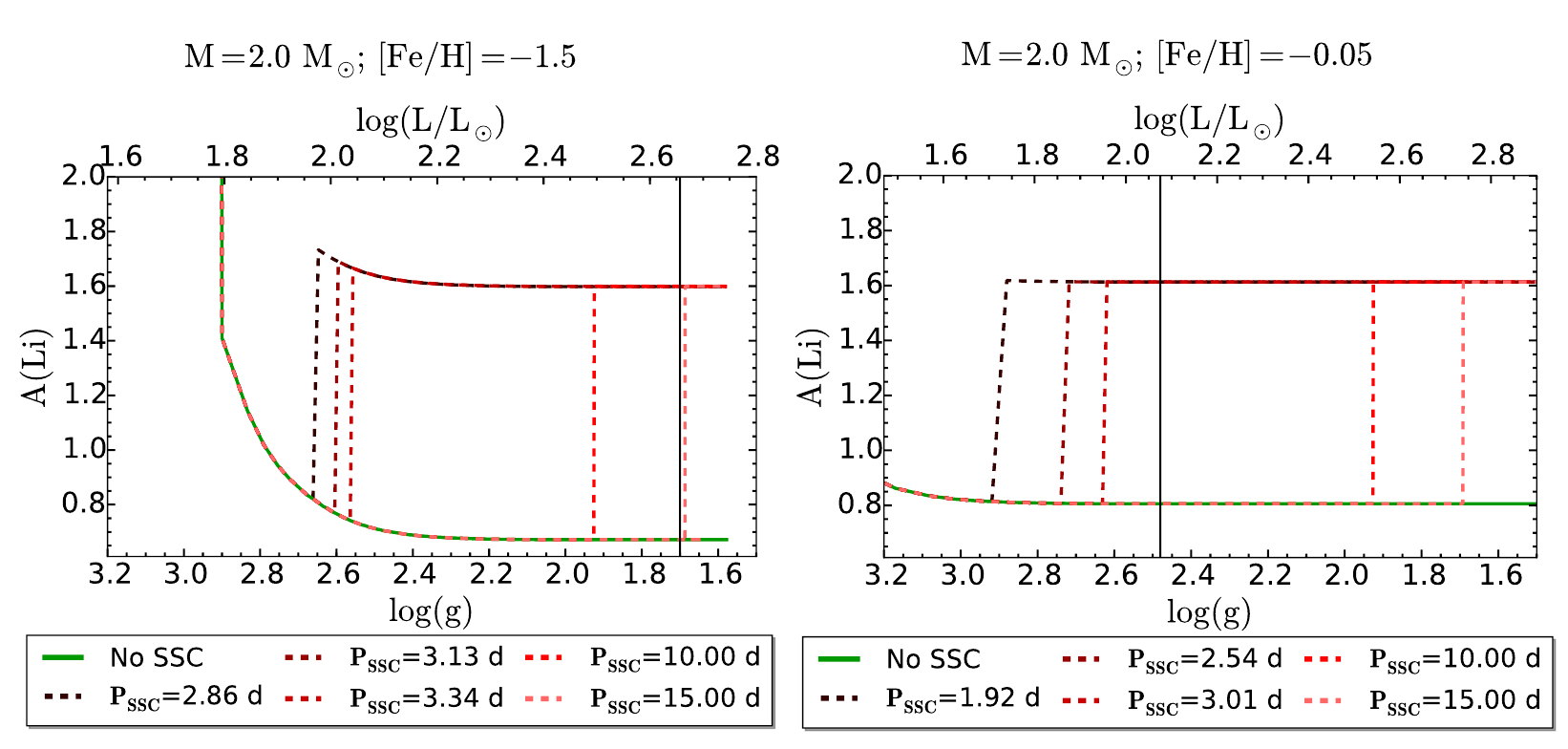}
\caption{Same as Figure \ref{Lisup} but for a $2\,\mathrm{M_{\odot}}$ host. We show only 2 different metallicities, because the trend is the same that for other masses, but the final $\mathrm{A(Li)}$ does not change depending on the period of the engulfed SSC.}
\label{Lisup20}
\end{center}
\end{figure*}

Figure \ref{Lisup} presents our models of engulfment of SSCs with different periods in the evolution of a $1.0$ solar mass host with $\mathrm{A(Li)_{*,ini}}=2.6$, for various metallicities. 
The green solid line represents the $\mathrm{A(Li)}$ evolution of the star evolving in isolation. Early on the RGB we can see the effect of the FDU, followed by an abundance plateau. A comparison between different metallicities shows that at lower metallicities stars have a higher Li abundance overall before the engulfment of the SSC, in agreement with what is known from standard stellar evolution: For the same mass, stars of lower metallicities have lower opacities, developing shallower convective envelopes during the RGB and thus presenting less mass for the same Li to dilute into.
Stars with 1.5 $M_{\odot}$ and 2.0 $M_{\odot}$ in Figures \ref{Lisup15} and \ref{Lisup20}, respectively, show the same trend of A(Li) with metallicity than 1.0 solar mass stars. Higher [Fe/H] stars have lower A(Li) in general, but the effect is less prominent with increasing mass. 

The accretion process produces in all cases a rapid increase in the Li abundance as soon as the content of the planet is mixed with the envelope. Note that, if the accretion occurs early enough on the RGB, one can also recognize the standard dilution due to the deepening of the envelope during the FDU. Given our post-processing approach, standard dependencies such as that of the depth of the FDU as function of metallicity, remain unaltered. 

It is clear that the Li abundance with SSC engulfment is always higher than that of the star evolving in isolation, but the exact superficial A(Li) post-accretion depends on all the stellar parameters. Stars with low [Fe/H] display the largest abundance after the accretion of the SSC given their initially higher abundances. Metal rich stars show the largest difference in A(Li) between the case in isolation and that with SSC engulfment. As an example, for the $1.0\,M_{\odot}$ star with [Fe/H]=-0.05, the $\mathrm{A(Li)}\sim -1.7$ without engulfment. With engulfment of a SSC with period $P_{SSC}=1.42$ days, the star reaches an abundance of $\mathrm{A(Li)}\sim1.9$, very high in comparison to its initial abundance.

Besides the dilution just discussed above, there can also be nuclear burning. The amount of planetary Li burned during the evolution of the star depends on the corresponding nuclear reaction rates. We will see a difference in the Li abundance as long as the element is burned under convective conditions. Sometimes, even if the temperature of the base of the convective envelope is over $T\sim2.5\times10^6$ K, there will be no burning in the star due to the lower densities found on the RGB phase.

For lower metallicities the Li is not burned in the convective envelope at any point in the evolution of the star, so we expect that SSCs of different periods shift the $\log(g)$ of the initial increase in Li abundance produced by incorporating the SSC Li into the star, but we do not expect any further changes in superficial abundance after that. For higher metallicity stars, there is burning of Li in the convective envelope during the FDU. The burning of planetary Li is not evident, as it is not seen as a sharp or noticeable drop off in A(Li) (Figure \ref{Lisup}), but some part of the SSC Li is being burned immediately after the engulfment, thus the maximum abundance obtained when the SSC is diluted is much less than the maximum abundance that is expected by just diluting all the SSC content in the convective envelope.

For stars of higher metallicities, there are small differences in A(Li) depending on the period of the engulfed SSC. As can be seen in the bottom right panel of Figure \ref{Lisup}, if the engulfment occurs before the end of the FDU, planets with smaller periods will produce a smaller enrichment in Li, because the rates of Li burning are larger (higher densities). After the maximum penetration of the convective layer there is no substantial burning of planetary Li due to the low densities of the region, and the A(Li) after this point remains almost constant, only decreasing progressively by a small amount.

It is important to notice that in our Figures \ref{Lisup}, \ref{Lisup15} and \ref{Lisup20}, the final Li abundance does not change for SSC accreted after the end of the FDU, as it can be seen for planets engulfed with larger periods. The low level of Li burning after this point in stellar evolution allows to place an upper limit in the A(Li) after engulfment by just analyzing any SSC accreted post-FDU.

The burning of planetary Li also produces slightly different A(Li) for various SSC periods (P in days) in the 1.5 $M_{\odot}$ stars at higher metallicity. In contrast, for higher mass stars the final Li abundance does not depend too much on the time of accretion, and is only dependent on the intrinsic properties of the SSC and star.

Up to this point, our exploration of parameter space has produced a maximum Li abundance of $\mathrm{A(Li)}\sim2.0$, after the accretion of a SSC of 15 $M_J$ by a $1.0\,\mathrm{M_{\odot}}$ star of [Fe/H]$=-2.0$. This represented an increase of $0.9$ dex over the original Li abundance of the isolated star. The increase is much larger for higher metallicity stars, even reaching a $4.0$ dex difference between stars with and without SSC engulfment (see the case of the $1.0\,\mathrm{M_{\odot}}$ star with [Fe/H] = -0.05 in Figure \ref{Lisup}), but never producing the most Li-rich giants, so this difference could not be measured, as the giants before the engulfment would only have upper limits in $\mathrm{A(Li)}$.

\subsection{Changing the engulfed SSC} \label{i_lissc}

Changing the intrinsic Li content of the SSC will produce different signatures in the star. In this section we model 4 realistic planets with different masses but also different Li abundances: a Jupiter mass planet with $Z_{SSC}=2.5\,\mathrm{Z_{\odot}}$, an Earth-like planet with rocky composition ($Z=1.0$), a brown dwarf of $15\,\mathrm{M_J}$ with Jupiter-like composition ($Z_{SSC}=2.5\,\mathrm{Z_{\odot}}$, slightly enhanced in metals in case it is formed by core accretion), and a brown dwarf of $15\,\mathrm{M_J}$ with solar metallicity. Higher mass SSCs are not considered because as it was seen in Section \ref{Pssc}, those objects would not dissolve in the convective envelope.

\begin{figure}[!htb]
\begin{center}
\includegraphics[width=0.5\textwidth]{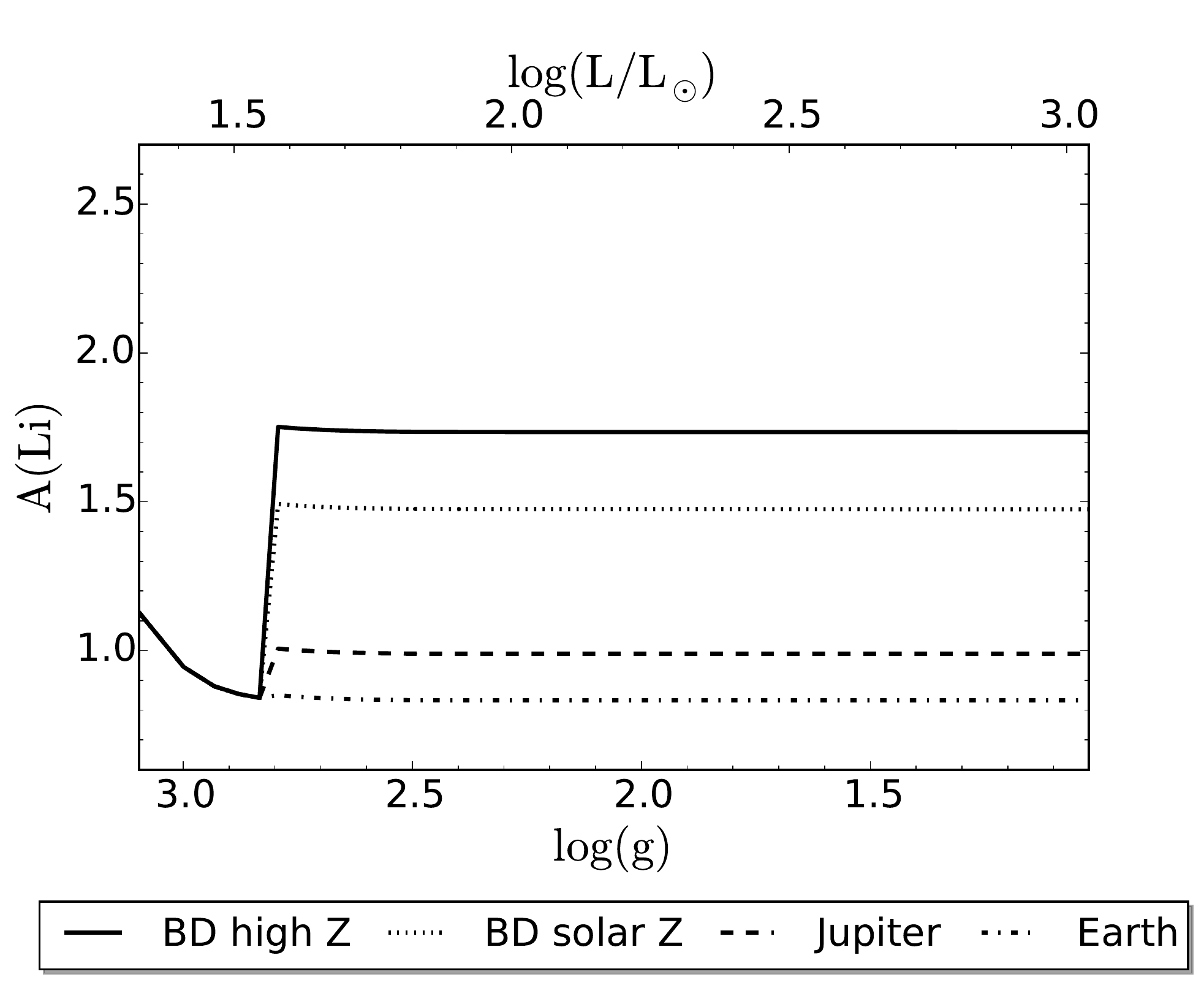}
\caption{Evolution of surface Li abundance in a $1.5\,M_{\odot}$ star with [Fe/H]=$-1.0$ that engulfs 4 different SSCs at the same timestep, a brown dwarf with metal enhancement, a brown dwarf with solar composition, a Jupiter-like planet and an Earth-like planet.}
\label{diffssc10}
\end{center}
\end{figure}

In Figure \ref{diffssc10}, we see a $1.5\,\mathrm{M_{\odot}}$ star of [Fe/H]$=-1.0$ engulfing four different SSC at the same timestep. One of these, for comparison, is the planet case studied in the previous section (the brown dwarf with metal enhancement), the one that produces the largest enhancement. A rocky SSC with the mass of the Earth produces almost no enhancement at all. Even if it has a larger metal content, the very low mass makes this the worst case scenario in terms of Li-enrichment. The BD with solar composition still produced a very large Li enhancement, only $\sim0.3$ dex lower than that of the BD with enhanced metal content. At last, a Jupiter-like planet will produce a signal that is $\sim0.2$ dex higher than that of the star.

\subsection{Changing the starting Li abundance of the host star} \label{i_lihost}

In this section, we will consider host stars with different starting Li abundances. We model stars with $\mathrm{A(Li)_{*,ini}}=3.3$, compared to giants with $\mathrm{A(Li)_{*,ini}}=2.6$ (as used so far). We will also model stars from the Li-dip, with a very small abundance $\mathrm{A(Li)_{*,ini}}\ll 1.0$, which covers all of the possibilities for MS stars in the literature \citep[for some catalogs of dwarfs with measured Li abundances see][and references therein]{ivan12, dm15}.

\begin{figure*}[!htb]
\begin{center}
\includegraphics[width=\textwidth]{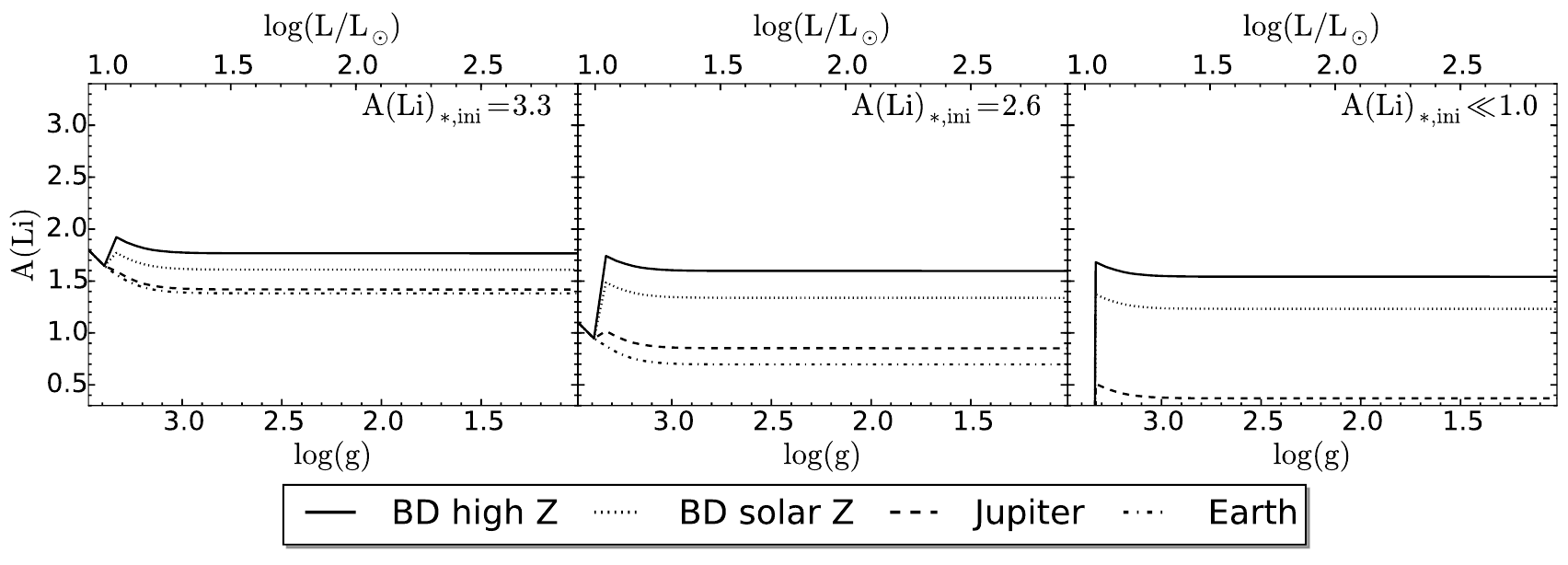}
\caption{Evolution of the superficial Li abundance for stars with the same mass and metallicity ($1.5\,M_{\odot}$ and [Fe/H]=$0.0$), but different initial Li abundances, as indicated on the different panels. These stars have engulfed a brown dwarf with metal enhancement, a brown dwarf with solar composition, a Jupiter-like planet and an Earth-like planet, all with an orbital period of $P_{SSC}=0.78$ days. In the right panel, the abundance of the star after the accretion of the Earth-like planet cannot be observed in the plot, but corresponds to a value of $\mathrm{A(Li)}=-0.7$.}
\label{diffstellar15_01879}
\end{center}
\end{figure*}

In Figure \ref{diffstellar15_01879} we show models for a 1.5 $M_{\odot}$ host star with [Fe/H]=$0.0$, where the panels represent stars with different initial Li abundances, from high initial Li abundance of  $\mathrm{A(Li)}_{*,ini}=3.3$ to low MS abundances  $\mathrm{A(Li)}_{*,ini}\ll1.0$. These stars are engulfing the same 4 planets discussed in Section \ref{i_lissc} (a metal enriched BD, a solar composition BD, a Jupiter and an Earth) with a period of $P_{SSC}=0.78$ days.

The trend with different SSCs is the same found on Section \ref{i_lissc}, where the BD with enhanced metal content produces the largest Li enrichment, and the Earth-like planet almost does not increase the Li abundance of the star. In the right panel, for the low MS abundance, the Earth-like planet produces a signature of $\mathrm{A(Li)}=-0.7$, that cannot be seen in the figure. Given the very low post-FDU value for these stars, even a low-mass planet like the Earth could produce an important increase in abundance, but that would still remain undetected, as even with engulfment the abundance is very low.

An interesting effect is that the larger the Li abundance of the host, the less difference there is in $\mathrm{A(Li)}$ between engulfment of different SSCs (focusing on one panel at a time in Figure \ref{diffstellar15_01879}). This is because if the star has a larger Li content, a larger amount of Li needs to be diluted to produce an observable signature. Then, if the Li content provided by the SSC is the same, it would produce a larger enhancement for the star with less Li during its isolated evolution.
This effect is similar to what we observed for stars of different metallicities in Figure \ref{Lisup}, where if the same SSC was engulfed, the star with a lower baseline of Li abundance will have a larger difference between A(Li) evolving in isolation and with engulfment. Also for similar reasons, if we compare the signature from the same SSC in different panels in Figure \ref{diffstellar15_01879}, we see that even if there is a large difference in Li content before the engulfment, the difference in Li abundances is always smaller after the accretion of SSC. The same planet dissolved in a region with more Li will produce a smaller signature.

The larger the Li content of the SSC, the smaller the difference in the signature when the companion is engulfed by stars of different $\mathrm{A(Li)}_{*,ini}$. The BD with enhanced metal content produces almost the same Li enrichment in the 3 different stars. When the mass of Li contributed by the companion is large, it will dominate the final Li abundance of the star, thus the Li of the star would only imply a small difference in the final signal.

Therefore the final abundance post-engulfment when the SSC produces a large enhancement is very similar between stars despite a large initial difference in the $\mathrm{A(Li)_{*,ini}}$.

\begin{figure*}[!htb]
\begin{center}
\includegraphics[width=\textwidth]{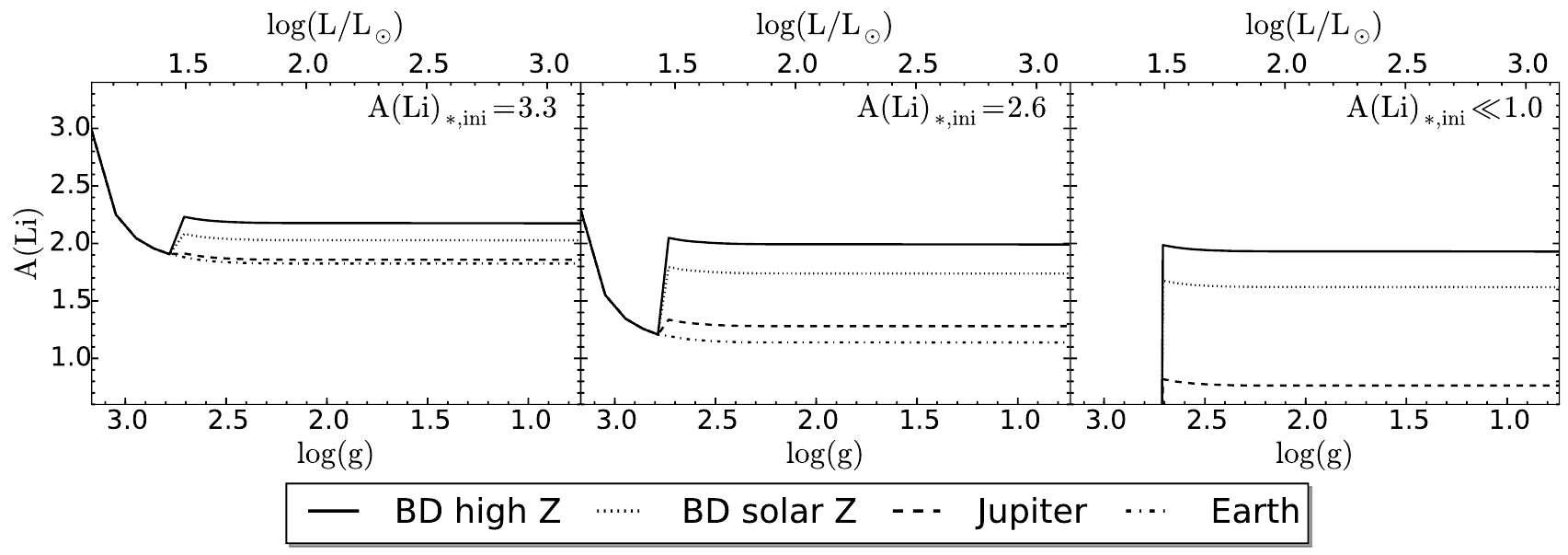}
\caption{Same as Figure \ref{diffstellar15_01879} but for host stars of $1.0\,M_{\odot}$ and [Fe/H]=$-2.0$ with different initial Li abundances, as indicated in the panels. These stars have engulfed a brown dwarf with metal enhancement, a brown dwarf with solar composition, a Jupiter-like planet and an Earth-like planet with a period of $P_{SSC}=2.00$ days. In the right panel, the abundance of the star after the engulfment of an Earth-like planet reaches a value of $\mathrm{A(Li)}\sim -0.3$.}
\label{diffstellar10_00020}
\end{center}
\end{figure*}

The same effects can be observed in Figure \ref{diffstellar10_00020}, but in a star of $1.0\,M_{\odot}$ and [Fe/H]=$-2.0$. We show this star in particular because this is the case where we obtain a larger Li abundance after engulfment of $\mathrm{A(Li)}=2.2$, when a star of high $\mathrm{A(Li)}_{*,ini}$ accretes a BD enhanced in metal content.

If 3 different mass domains are considered, with low-mass stars having an abundance of $\mathrm{A(Li)_{*,ini}}=2.6$, intermediate mass stars that have evolved from the Li-dip, and as such have $\mathrm{A(Li)_{*,ini}}\ll 1.0$ and higher-mass stars that have a higher initial abundance of $\mathrm{A(Li)_{*,ini}}=3.3$ (See Section \ref{starsec} for more details), then we can explore the SSC masses that would produce a clearly identifiable signature from those of giants in isolation as function of stellar mass.
We set a threshold in Li abundance for each mass range and identify which SSCs would produce such a signature for [Fe/H]$=-2.0$ to [Fe/H]$=0.0$. For low and high mass stars, the threshold is identified as the highest $\mathrm{A(Li)}$ of giants evolving in isolation. In this case, a higher signal could only be produce by the engulfment of a companion. For the intermediate mass giants that have evolved from the Li-dip, we choose 2 different thresholds of $\mathrm{A(Li)}=0.5$ and $\mathrm{A(Li)}=1.0$, as any planet could be able to produce a signal higher than that of stars evolving in isolation, given their very low abundances. In Table \ref{detect} we can see the minimum planetary mass to produce a given signal.

\begin{table*}[!htb]
\begin{center}
\caption{Limiting SSC mass to produce an engulfment signature that could not be mistaken with isolated giants in three different stellar mass ranges.\label{detect}}
\begin{tabular}{ccc}
\tableline\tableline
Mass range & $\mathrm{A(Li)}$ threshold & Planet mass ($\mathrm{M_J}$)\\

\tableline
Low-mass stars & $1.25$ & $6.7$ \\
Evolved Li-dip stars &$0.5$ &$2.6$ \\
Evolved Li-dip stars &$1.0$ &$8.6$ \\
High-mass stars &$1.52$ &$6.0$ \\
\tableline
\end{tabular}
\end{center}
\end{table*}

Planets with smaller masses may still produce a giant with enhanced Li abundance, but the values quoted in Table \ref{detect} show the minimum planetary mass that would not be confused with the signal of stars evolving in isolation, and is always produced by SSC engulfment.

We have proven so far that the effect of SSC engulfment in Li abundance of the stars is linear, thus, if more than one SSC is accreted they would produce the same effect as a companion with the combined mass of those smaller SSCs.
\subsection{Rates of Li-enrichment}\label{rates}

Using distributions for planet periods and masses, and extrapolating those up to $15\,\mathrm{M_J}$, we can predict the rate of Li-enrichment in giants exclusively by SSC engulfment. We consider SSCs only in this mass range because higher mass objects would no produce an enhancement, as discussed in Section \ref{Pssc}.

The planetary mass distribution we are using follows that of \citet{udry07}, with a steep rise towards the low-mass planets. The distribution in \citet{how10}, that extends to even lower planetary masses, is also used in some of our models to test if the use of a different distribution would change the Li abundance distribution, but these very small planets produce a very low signature of Li-enrichment in the star. Thus, using the two different occurrence rates only changes the final Li distribution by a negligible amount. We have extended these distributions to include low mass brown dwarfs (or high mass planets) of up to $15\,\mathrm{M_J}$. 

\begin{figure*}[!htb]
\begin{center}
\includegraphics[width=0.7\textwidth]{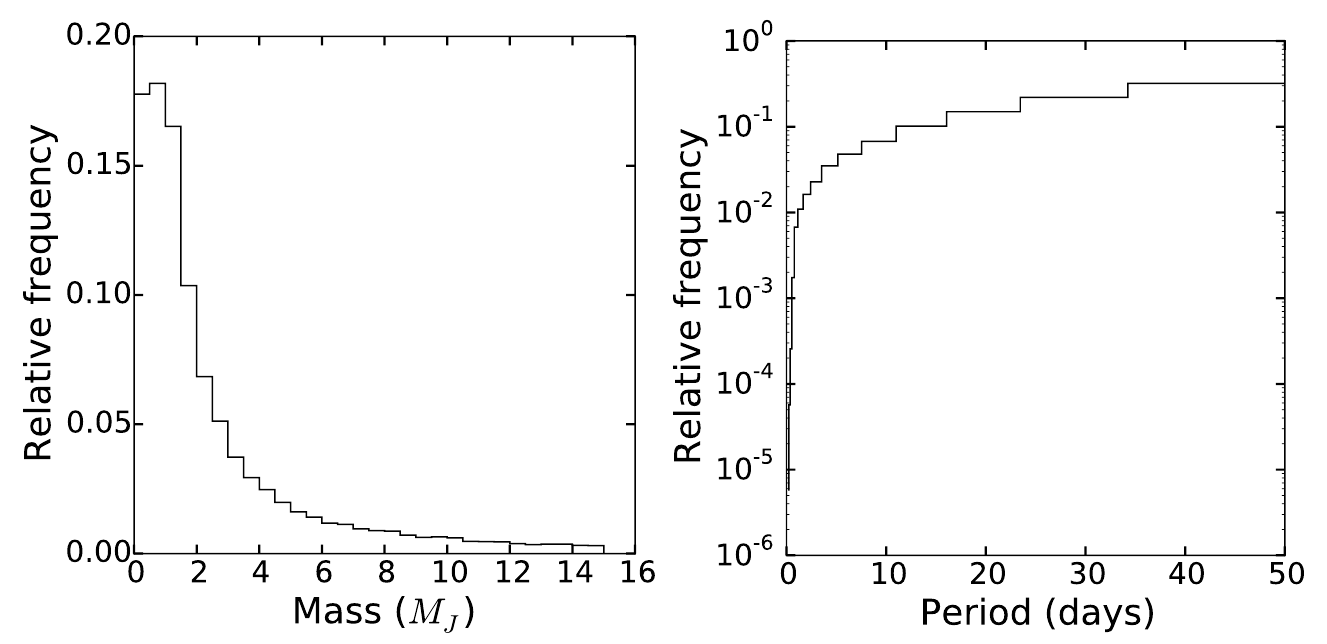}
\caption{Distribution of planet masses (left panel) and periods (right panel) used for our simulated population.}
\label{dis_planets}
\end{center}
\end{figure*}

The results presented throughout this section consider a period distribution from \citet{how12} for small planets ($2-4\,M_{\oplus}$) with periods $P<50$ days. We also use their period distribution for giant planets ($8-32\,M_{\oplus}$) consistent with the distribution from \citet{cum08} that also considers planets with periods up to $P=2000$ days for the gaseous planets and brown dwarfs. We show the final period and mass distribution for the simulated SSCs in Figure \ref{dis_planets}.

We generate a synthetic population of stars with masses between 1.0 and 2.0 $\mathrm{M_{\odot}}$, and metallicities between [Fe/H]=$-2.0$ and [Fe/H]=$0.0$, following the metallicity distribution of known Li-rich giants, with a large increase in number towards metallicities closer to solar. These stars engulf companions of a given period and mass which follow the distributions indicated previously. Then, our grid of models of Li abundances is interpolated to obtain the $\mathrm{A(Li)}$ after accretion of SSC.

The initial Li abundance of the star is chosen according to its mass, as described in Section \ref{starsec}. Low mass, old halo stars will have an initial abundance of $\mathrm{A(Li)}_{*,ini}=2.6$, intermediate mass stars that have evolved from the Li-dip will have $\mathrm{A(Li)}_{*,ini}\ll1.0$ and high mass stars will have a higher Li abundance of $\mathrm{A(Li)}_{*,ini}=3.3$ (corresponding to stars with ages$<2.0$ Gyrs). By considering these 3 different mass domains we are also taking into account the difference in ages of stars and their corresponding abundances.

\begin{figure*}[!htb]
\begin{center}
\includegraphics[width=0.95\textwidth]{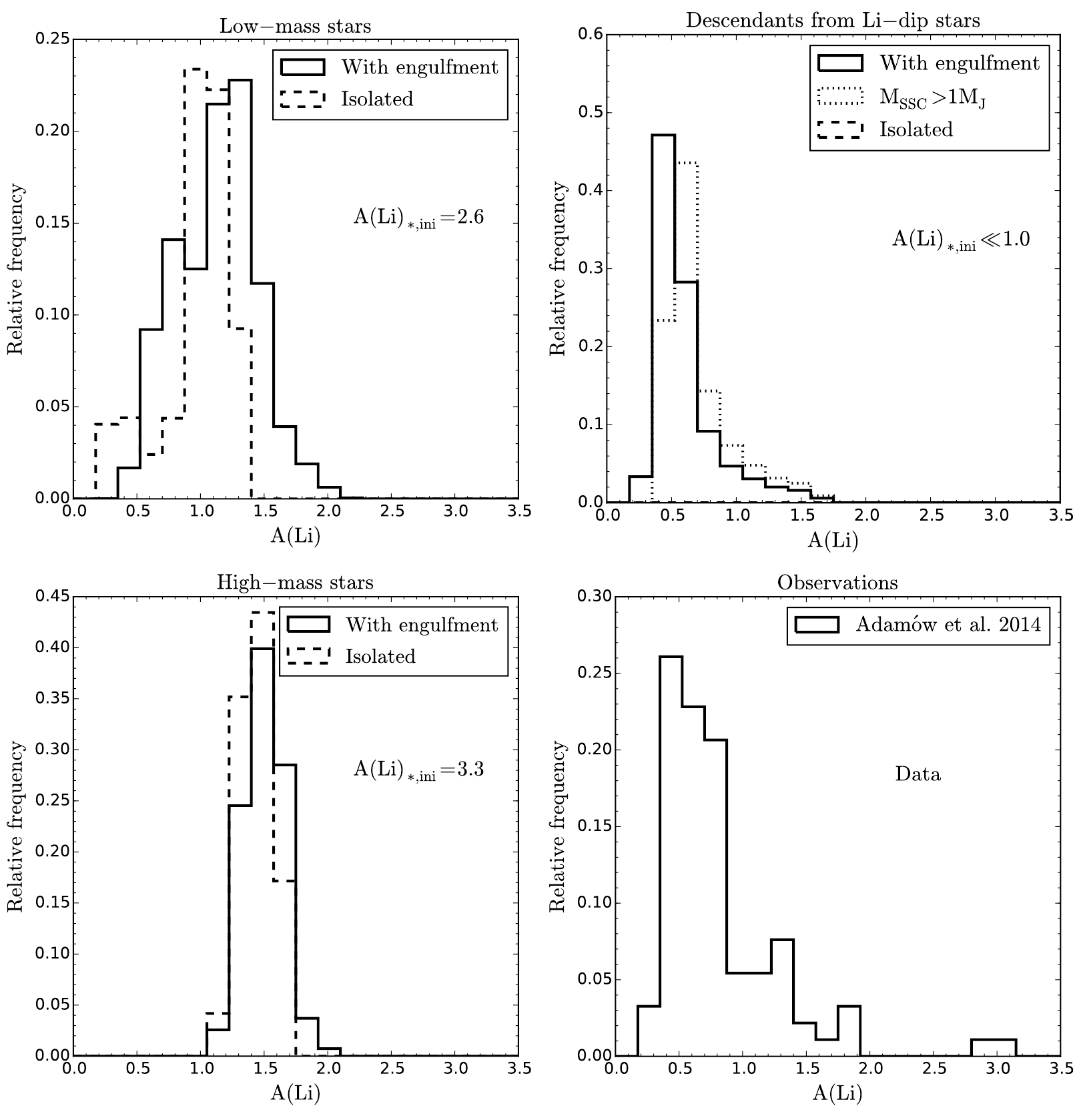}
\caption{Li abundance distribution of giants obtained by using our models of SSC engulfment up to the RGB bump, for the case when all stars engulf their planet, compared to a population of stars evolving in isolation. The distributions consider three different mass domains as indicated in each panel and in the bottom right panel, the distribution of Li abundances for field stars of \citet{ada14} is shown. For high mass stars, there is little difference between the population evolving in isolation and that with planet accretion. For stars that have evolved from the Li-dip, the abundance in MS is so low that stars evolving in isolation would only have upper limits, and as such we cannot see their distribution in this Figure. We also compare the distribution with one produced by the engulfment of SSCs only with masses $M>1.0\,\mathrm{M_J}$. For low mass stars, the peak of the distribution moves towards higher Li, with a gradual decrease towards higher $\mathrm{A(Li)}$.}
\label{DistLi}
\end{center}
\end{figure*}

Figure \ref{DistLi} shows the distribution of $\mathrm{A(Li)}$ in simulated populations of 35000 stars after the engulfment of their SSCs, where all stars had initially a SSC. Each of the panels represents a population with a different mass range, as indicated in each panel. For reference, the distributions in each case are compared to that of a synthetic population that does not go through the SSC accretion process. The Li abundance of these stars was evaluated at $\log(g)=1.5$.

Although for low and intermediate mass stars all of the giants have abundances lower than $\mathrm{A(Li)}=1.5$, as expected from canonical stellar evolution, the higher mass stars could have an abundance higher than that even with no accretion. These, which would be considered Li-rich giants by usual criteria, are just ``normal" stars.

For low mass stars, there is a clear increase of giants with higher Li abundances, showing Li-rich giants with $\mathrm{A(Li)}=1.5$ to $\mathrm{A(Li)}=2.2$ that the population not experiencing engulfment does not show. The peak of the distribution for giants with engulfment is shifted towards higher abundances when compared to the isolated stars, and the distribution presents a gradual decrease from the peak towards higher abundances.

The intermediate mass stars that have left the MS with very low Li abundance, have only upper limits in their giant evolution, thus, the distribution of isolated stars cannot be seen in the upper right panel of Figure \ref{DistLi}. The location of the peak of this distribution depends on the assumptions of the abundance of Li-dip stars and it does not change the abundance post-engulfment. Stars in this mass range would produce a clean sample where the engulfment scenario could be tested, as there is no contamination from giant stars that have evolved in isolation, thus emphasizing the importance of this subject of reliable determinations of mass and evolutionary stages of stars. The peak of the distribution of intermediate-mass stars that have engulfed a SSC is centered at $\mathrm{A(Li)}=0.5$, and barely reaches the abundances of Li-rich giants as typically defined observationally, even if these are in reality Li-enriched giants produced by SSC accretion. In the upper right panel we also compare with the distribution of Li abundance produced by the engulfment of SSCs with masses $M>1.0\,\mathrm{M_J}$. It is very similar to the distribution including the engulfment of all SSCs masses, but the peak of the distribution  moves towards higher abundances. This means that even smaller mass planets have an impact on the final distribution of these giants that have evolved from the Li-dip.

The distribution of high mass stars evolving in isolation looks very similar to giants that have accreted their SSCs. In this population there would be almost no sign of accretion even if this process is actually changing the abundance of the stars.

The data shown in the bottom right panel corresponds to that in \citet{ada14}, since this work provides not only information about the Li-rich giants, but of the complete sample of stars, including those with lower abundances and considering either stars in the red giant branch or in the clump, that may have experienced additional depletion that takes place during the RGB evolution after the luminosity function bump. This is the most interesting comparison available with field giants. Since these field stars are not only located in a specific mass range, to properly compare our simulated distribution with observations, a combination of the three different mass ranges would be needed, and we would have to assume a fraction of giants that engulf their SSC. Also, we are aware that the assumptions used here about the initial Li abundance of each mass range are oversimplifying the problem. Given that there is some MS depletion, the peaks of the synthetic distributions would most likely move towards lower Li abundances. The natural extension of our simulations is to consider the Li abundance at turn-off of different clusters and its dependence on mass and metallicity and simulate giants according to that.
Although SSC engulfment can explain the presence of stars in the high Li abundance end of the observed distribution, it could clearly not explain the peak observed close to $\mathrm{A(Li)}=3.0$. These giants would require a different source of Li.

It is relevant to note that if the brown dwarf population is much different than what we considered in our models (as it could be the case if the brown dwarf desert is real), the higher Li abundance end of the rates calculated here would change, but high mass planets would still produce Li-rich giants.

An interesting future test could be to check the Li abundance as a function of age of giants with Gaia parallaxes plus seismology from TESS, which would have a known evolutionary stage and mass, and as such could be placed in one of the specific populations modeled. The same could be done with Kepler giants with spectroscopic data from APOGEE.

We would like to emphasize that, as seen from our models, the currently used definition of Li rich giants ($\mathrm{A(Li)>1.5}$) leaves out some giants whose Li abundance cannot be explained by canonical evolution without engulfment. Moreover, high-mass stars that do not go through any unusual process could have higher abundances than the standard limit for Li-rich giants used. Thus, low mass and intermediate mass giants that have accreted SSCs, may look as normal evolved giants, while high mass stars evolving in isolation may look as Li-rich even if they are just ``normal" red giants.

To exemplify this, we combine the fraction of giants that end up being Li-rich according to the traditional definition of Li-rich giants after our modeling of the engulfment process, and the fraction of stars that have planets in close orbits during the MS, which may be eventually engulfed when the star becomes a red giant. We get a total rate of Li-enrichment ($\mathrm{A(Li)>1.5}$) by SSC engulfment in the different mass ranges, ignoring all dynamical effects.
For close-in planet fractions of $10\%$ and $30\%$ during the MS, low-mass stars could produce from $1.0\%$ to $3.0\%$ of giants with abundances higher than $\mathrm{A(Li)}=1.5$, very similar to the fraction reported in the literature. For intermediate mass stars that have evolved from the Li-dip it would seem that they produce a much smaller rate of enhancement, from $0.2\%$ to $0.6\%$, but clearly this result is misleading, as although the fraction is lower according to the limit $\mathrm{A(Li)}=1.5$, we know that as every isolated giant has only an upper limit, any planet would produce a signature in the giant, and would be producing an unusually enriched star. Then, the relevance of this population is not the high contribution to giants with very high abundances, but the very high signal, since every detection would be a giant going through engulfment and Li-enrichment. The Li-dip evolved stars then is a clean sample to test this scenario. Once again, to exemplify that using the common definition of Li-rich giants produces misleading results in the rates, we see that high mass stars produce a fraction of Li-rich giants of $3.4\%$ to $9.6\%$, much higher than in the low mass case, as a considerable number of giants evolving in isolation in this mass range would already have high Li-abundances.

\section{Discussion}

\begin{figure}[!htb]
\begin{center}
\includegraphics[width=0.45\textwidth]{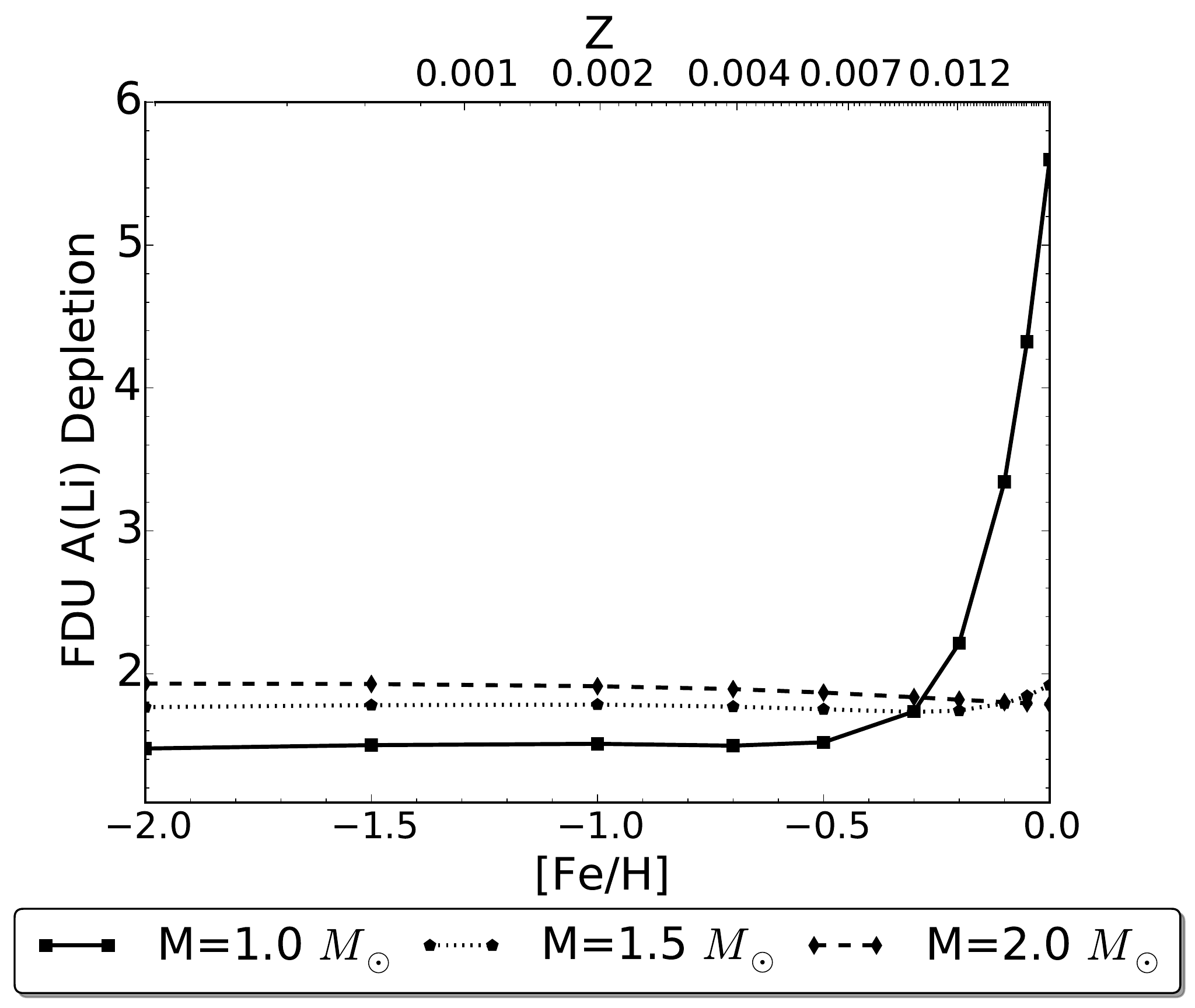}
\caption{Li abundance depletion factors as a function of metallicity, that compare the abundance of stars evolving in isolation after the FDU with abundances in MS. Different stellar masses are represented by different symbols and linestyles, as indicated. We can see a large depletion in low mass metal rich stars that is produce by the combined effect of dilution and burning.}
\label{AlivsZ}
\end{center}
\end{figure}

To put the models of SSC engulfment in perspective, we present in Figure \ref{AlivsZ} the Li abundance dilution factors of stars of different masses and metallicities evolving in isolation
\begin{equation}
\mathrm{Depletion}=\mathrm{A(Li)_{MS}-A(Li)_{post\,FDU}}
\end{equation}
This means that a depletion factor of $1.0$ corresponds to a factor of ten reduction of Li during the FDU.
As we can see here, although the FDU is weakly mass and metallicity dependent in most cases, for $1.0\,\mathrm{M_\odot}$ stars of high metallicity, we see a very strong depletion that is not only due to the dilution because of the deepening of the convective layer, but also because Li is burned in the bottom of the convective zone on the approach to the giant branch.

This indicates that metal rich stars of low masses should not show any Li according to canonical stellar evolution. As such, this could be an interesting sample to study.
When these stars engulf a SSCs with a large mass of Li (as the $15\,\mathrm{M_J}$ SSC), its final abundance post-engulfment is dominated by the companion, an effect already discussed in Section \ref{i_lihost} when comparing stars with different initial Li abundance, so despite the large dilution factors, the abundance after accretion of these large companions would be the same independent of metallicity for stars of the same mass.

\begin{figure}[!htb]
\begin{center}
\includegraphics[width=0.45\textwidth]{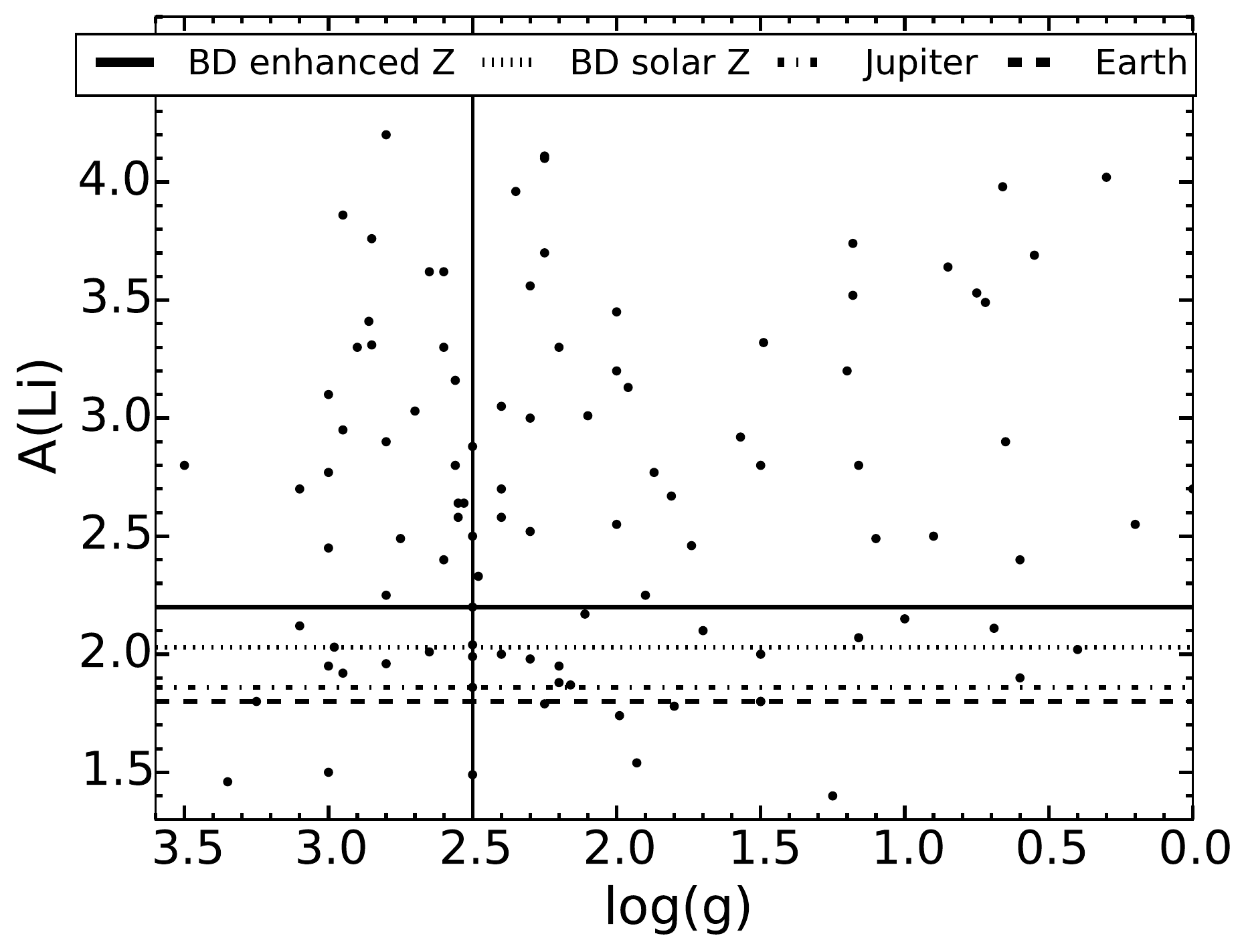}
\caption{Observations of Li rich giants compared to the maximum obtained A(Li) from models of engulfment with SSC with four different masses indicated by the horizontal lines. The vertical line shows the location of the luminosity function bump for the $1.0\,\mathrm{M_\odot}$ of [Fe/H]=0.0. Other modeled stars have their RGB bumps at lower $\log(g)$ values. Here it is evident that most of the giants with smaller Li abundances could be explained by SSC engulfment, but the giants with higher A(Li) need a different source. The observations of Li-rich giants are obtained from \citet{bro89, lh07, ruch11,kum11, kir12,ms13,ada14}, and references therein.}
\label{CompObs}
\end{center}
\end{figure}

We compare directly our models with some observations of Li-rich giants (Figure \ref{CompObs}). The horizontal lines represent the maximum abundance that can be obtained by the engulfment of a different SSC and we show as a vertical solid line the location of the luminosity function bump for the $1.0\,\mathrm{M_\odot}$ star of [Fe/H]=0.0. The RGB bump for the rest of the modeled stars lies towards lower $\log(g)$ values. 

After exploring the parameter space we can identify the properties that affect the most the final Li abundance. Changes in metallicity would produce a variation of at most $0.5$ dex, and the orbital period of the SSC does not affect the final Li abundance of the star, making the mass and metal content of the SSC the most important planetary parameters involved.

The highest Li abundance reached at the red giant phase from SSC engulfment is lower than $\mathrm{A(Li)}=2.2$, and thus the Li-rich giants with abundances higher than this limit must be explained by some other mechanism. As the engulfment of a SSC seems to be one of the few known mechanisms for Li enrichment capable of working before the RGB bump, Li-rich giants in this evolutionary phase with $1.5\leq\mathrm{A(Li)}\leq2.2$ are most likely to be formed by this channel. 

Our results agree with order of magnitude calculations done in other works that explore planet engulfment. \citet{jol13} warn that one must be realistic about the amount of Li that can be obtained from such process, as do others that have found specific Li-rich giants and study if those cases can be explained by accretion \citep{dor15,jol15}. Our work quantifies the impact of planet properties, mass and composition, which improves over rough estimates from individual objects.

The calculated rates of Li-enrichment in the lower-mass range also agree with observations when choosing the commonly used definition of Li-rich giants. We see that from $1\%$ to $3\%$ of the giants should be Li-rich when produced by engulfment in our models, while observations report a value between $1\%$ to $2\%$. However, rates calculated by considering enrichment only when $\mathrm{A(Li)\gtrsim1.5}$ are misleading. We can see this for giants evolved from the Li-dip: in this mass range, every engulfment process produces a Li-enriched giant, so the fraction of enrichment should be very high. Instead, when applying the traditional limit, we only find rates of enrichment of $0.2\%$ to $0.6\%$. The same occurs for high mass stars, where we obtain a very high rate of enrichment produced mainly because isolated giants have high abundances, and not because planet engulfment is increasing the amount of Li-rich giants.

The predictions of Li abundance in populations engulfing planets obtained in Section \ref{rates} should be extremely useful to compare with observations of stars before the luminosity function bump where there is no contamination from giants that can obtain their Li in some other way. One other sample with which we could test our predictions are those stars belonging to the Li-dip. The distribution of stars that have evolved from the MS Li-dip shows how a sample of stars undergoing SSC engulfment would look like in a relatively clean sample with no other effects on the abundance, as isolated giants would only have upper limits. If only some stars in those samples have measurements of Li abundance instead of upper limits, those would be clear indications of this process and would even allow to test the underlying planetary distributions in a stellar mass range that is harder to explore using radial velocities. 
In contrast, high mass stars would not be a good population to test the accretion of SSCs by giants, as the sample with and without engulfment produce a very similar distribution in the Li abundance.

\subsection{Other observational constraints}

Some observational evidence has been used to deem the planet ingestion scenario unlikely. Other light elements, like beryllium have been used to distinguish between Li replenishment scenarios. In the work by \citet{melo05}, the authors interpret the lack of $\mathrm{{}^9Be}$ as an indicator of internal production mechanisms, arguing that planet engulfment would have to increase both Li and beryllium at the same time, which is not observed.
In our models we also include beryllium in the calculations. We assume that the ingested $15\,\mathrm{M_J}$ brown dwarf has meteoritic composition ($\mathrm{A(Be)}$=1.42, \citet{lod98}), and find that in the best case scenario, $\mathrm{A(Be)}=0.7$ post ingestion, with a post-FDU abundance of the giant of $\mathrm{A(Be)}=0.6$, increasing very little the amount of stellar beryllium. All of the giants analyzed in \citet{melo05}, even those without high Li abundance, have extremely depleted beryllium. If an abundance of $\mathrm{A(Be)}=-5.0$ is considered for the post-FDU abundance of the star, the engulfment of this SSC would produce a signature of $\mathrm{A(Be)}=-0.22$, that according to \citet{melo05} should be detectable. If in fact the abundance of giants is as low as they say, any SSC could produce a measurable signature and the lack of measurements in Li-rich giants would indicate, in this case, that these are produced by other mechanism. Otherwise, it seems that SSC engulfment produces a small signature in the star, that could be very similar to the abundances found for isolated red giants, and thus may not be able to provide too much information on the SSC ingestion.

Another piece of information that could be used against the planet engulfment scenario is the evolutionary state of the stars. An initially promising, although mistaken idea was that Li-rich giants were associated with the luminosity function bump. Using Hipparcos parallaxes and evolutionary models, \citet{cb00} found that their limited sample of Li rich stars were most-likely located at this evolutionary phase, speculating that because of the destruction of the chemical composition barrier, the Cameron-Fowler mechanism could be acting and the rapid mixing necessary to keep Li from burning could proceed. Since then, there seemed to be an association between the Li rich giants and the luminosity function bump, and as such, models concentrated specifically in explaining giants after the RGB bump \citep[e.g.][]{dh04}. This was used against the planet engulfment scenario, since there is no reason why SSC accretion may have been taking place in an specific region of the color magnitude diagram. Currently, the detection of Li-rich stars not located at the red giant branch bump contradicts these earlier models.

More recent observations show that it seems there is no clear correlation of the Li rich giants with a particular phase of evolution \citep{gon09, leb12}, and there is also a large amount of Li-rich giants located around the clump \citep{kum11}. If this is the case, an enrichment episode could be triggered during the He-flash for low mass stars. \citet{den12} has also proposed an alternative scenario to explain the position of those stars in the color-magnitude diagram. According to that picture, those clump stars are actually closer to the RGB bump, but due to extra-mixing inside the star, they make an excursion in the HR-diagram, being located near the clump. If there is not a well defined phase of Li enrichment, it could mean that the process producing the enrichment is stochastic. An additional complication is that most of the observed Li-rich giants are in a region of the color-magnitude diagram where they could as easily be a low mass first ascending RGB stars as a more massive clump star, so even if the discussion may persist it is not entirely conclusive. This, along with our results that depending on the mass range, the most commonly used definition of a Li-rich giants might be mistaken, as isolated massive stars could have higher Li abundances while giants engulfing a SSC could have a lower abundances (See Section \ref{rates} for more details on this discussion), are some of the reasons why obtaining precise measurements of evolutionary stage and mass of giants is so important in finding the mechanism behind the enrichment. Also, some Li-rich giants could simply have not mixed their material yet (if they are caught before the FDU) or could actually be more massive stars that do not experience this type of mixing at all on the RGB phase. Evolutionary phase determination is uncertain, but the use of techniques as asteroseismology in Li-rich giants \citep{sa14, jof15} and others that help to precisely measure ages and distances are extremely helpful to confirm the mass of the stars and constrain the enrichment mechanism of the giants.

In the same context of distinguishing between SSC accretion scenario and internal Li production in the star, other observational signatures like rotation are used. Red giants are usually slow rotators, as a result of angular momentum conservation when their radii expands. In consequence, observed rapidly rotating red giants are also unusual and not easily explained. Planet or brown dwarf engulfment could be a solution for both the enhanced rotation and enriched Li abundance in red giants, where the external body is not only providing material, but is also a source of angular momentum \citep{jol09}. Then finding stars with both anomalies is a good indication of recent SSC engulfment, since the rotation signal will fade when the giants expand and slow down. \citet{dra02} found that a large portion of rapid rotators are Li rich giants, results confirmed by \citet{jol12} with a more uniform sample, that also pointed out a trend where rapid rotators are on general more Li-rich than rotationally normal RGB stars. This is not in conflict with our results. On the contrary, SSC engulfment can clearly increase the Li content of the star, but stars in those samples with a higher abundance than $\mathrm{A(Li)}\sim2.2$ should probably be treated with caution as they are most likely formed by a different mechanism, or more than one mechanism could operate.

As there is a possible association between Li-rich and rapidly rotating giants, we compare our fraction of enrichment with the fraction of rapid rotation from \citet{jol09}. The mass average of their sample of planet hosting stars is $M=1.1\,\mathrm{M_{\odot}}$, that corresponds to our low-mass simulation. Under some assumptions about the RGB lifetime, the fraction of stars with planetary companions and the fraction of planets that could make the star rotate faster, they obtain fractions of rapid rotators produced by engulfment from $0.53\%$ to $0.58\%$. This is lower than the fraction $1\%$ of Li-rich giants produced by planet ingestion in our low mass sample, although some of our assumptions (as the number of stars that host planets during the MS) are different. It could be expected that the fraction of rapid rotators produced by engulfment is lower, since even if the Li abundance could be conserved during the RGB evolution of the stars, the rotation of these stars slow down, so some of the Li-rich giants produce would not longer be rapidly rotating. This could also imply that the ingestion of some SSCs which can increase the Li abundance may be unable to spin-up the star.

The carbon isotopic ratio ($\mathrm{{}^{12}C/^{13}C}$) is another tool to identify mechanisms, that was not explored in our models. The FDU not only dilutes Li, but also decreases the $\mathrm{{}^{12}C/^{13}C}$. As $\mathrm{{}^{13}C}$ accumulates near the hydrogen burning shell, extra mixing reduces the carbon isotopic ratio. In contrast, engulfment of a SSC would not dramatically impact the surface abundance. \citet{jol12} measured $\mathrm{{}^{12}C/^{13}C}$ for their sample of stars and for the \citet{kum11} sample. While theirs show a ratio consistent with planet accretion, some stars at \citet{kum11} show reduced $\mathrm{{}^{12}C/^{13}C}$, a sign for extra mixing. This is already pointing out, from an observational point of view, that Li-rich giants may be produced by multiple mechanisms. 
On the other hand, while studying rapidly rotating giants, \citet{tay15} obtain the carbon isotopic ratio of their sample and as it is relatively high, they consider their rapidly rotating sample to be more likely the result from winds from evolved giants, but it could also be the result of planet accretion. Li measurements, not included in APOGEE, would be needed to confirm this. 

Even if SSC accretion could not explain the observations by the dilution of planetary Li content into the star in all cases, this does not mean that the presence of SSCs is not important for Li enrichment in giants, as suggested by \citet{ada14}. In recent studies on open clusters \citep{dm16, Jolarx} it has been concluded that red giant planet hosts seem to have a higher A(Li), but it is not known if that is related directly to the presence of a SSC and how. It is possible that accreting material from a SSC triggers some mechanism of extra-mixing allowing to bring processed material into the convective layer and thus increasing the Li content of the star, and thus increasing the limiting Li abundance we find only due to the dilution of the SSC of $\mathrm{A(Li)}=2.2$. SSC engulfment could therefore act as an indirect mixing trigger for Li production above the RGB bump. This would be difficult to distinguish from stellar mass transfer, however. We therefore focus on lower luminosity giants where mixing is suppressed.

\section{Summary}
The origin of Li rich giants is still a puzzle to this day. The different scenarios that could produce an enhancement (Li production inside the star plus an extra-mixing mechanism, accretion from an AGB star, and SSC engulfment) produce different observable signatures, and may even produce different Li enrichment. The different processes producing the Li-enrichment could also be generating Li-rich giants in different evolutionary stages, or different mass and metallicity domains.

To test the companion engulfment scenario, we have modeled the accretion of a SSC by low mass giants of different masses and metallicities, using reasonable assumptions on the population of SSCs. This provides upper limits on the lithium content during the RGB phase of stars after engulfment and allows to predict the maximum Li abundance of stars before the first dredge up, i.e., in those domains where extra-mixing is not efficient and could not produce Li-rich giants.

We calculate the point of SSC dissipation and find that objects with higher masses than $M_{SSC}>15\,\mathrm{M_J}$ dissolve in the radiative interior and cannot produce an observable signature in the star. Thus, objects with $15\,\mathrm{M_J}$ are the most efficient source of Li for giants. Objects with smaller masses can increase the superficial Li abundance of the star up to $\mathrm{A(Li)}=2.2$. The way in which our models are created produces a trustworthy upper limit, so any giant with a higher abundance than this is not produced by SSC engulfment. Another mechanism, like Li production plus additional non-canonical physics, or mass accretion from an enriched AGB star, will be needed to explain higher Li abundances.

The largest enhancement is always produced by a low mass brown dwarf with a metal content of $2.5\,\mathrm{Z_{\odot}}$, while an Earth-like planet almost does not increase the abundance of the star. Changing the initial Li abundance of the star almost does not change the Li abundance post engulfment for the SSC with higher Li content (as the metal enhanced BD), but it can produce large variations in Li when a lower Li content object is accreted.
We have seen in our models that the process of Li enrichment by SSC engulfment is linear, thus, the accretion of more than one companion is equivalent to the accretion of a more massive object with the combined mass of the smaller companions, as long as these have a mass smaller than $\sim15\,\mathrm{M_J}$, or otherwise they would dissolve in the radiative interior of the star.

We have calculated a limit to the mass of the ingested planet that could be detected in three different stellar mass ranges. Low mass stars ($1.0\leq\mathrm{M}<1.3\mathrm{M_{\odot}}$) with an initial abundance of $\mathrm{A(Li)}_{*,ini}=2.6$ that engulf planets with masses $M>6.7\,\mathrm{M_J}$ can be clearly identified from isolated stars. In this mass range, the fraction of giants with abundances $\mathrm{A(Li)}\gtrsim 1.5$ represent a $10\%$ of stars engulfing a SSC, implying a rate from $1\%$ to $3.\%$ of all giants. 

In the intermediate mass range, with giants that have evolved from the Li-dip ($\mathrm{M}\sim1.3\mathrm{M_{\odot}}$), with an abundance $\mathrm{A(Li)}_{*,ini}\ll1.0$, we find one of the most interesting populations to test the SSC engulfment scenario, since even lower mass objects engulfed could produce a Li abundance higher than those of stars in isolation. To produce giants with abundances $\mathrm{A(Li)}\gtrsim 0.5$ a $M_{SSC}>2.6\,\mathrm{M_J}$ is required. The fraction of giants with abundances $\mathrm{A(Li)}\gtrsim 1.5$ in this mass range is only $2\%$ of the giants that accrete a SSC, producing $0.2\%$ to $0.6\%$ of Li-rich giants by the commonly used criterion, when in reality this fraction is significantly larger, as any produced detection in this mass range would be evidence for engulfment.

The high mass stars (up to $2.0\,\mathrm{M_{\odot}}$) have a much higher abundance of $\mathrm{A(Li)}_{*,ini}=3.3$ initially, and consequently they have the highest abundances post-FDU of giants evolving in isolation. To identify the signature from engulfment without a possible confusion with these isolated giants, planets of masses $M_{SSC}>6.0\,\mathrm{M_J}$ must be consumed by the star. The larger Li abundances for these stars in isolation, when seen under the commonly-used criterion of $\mathrm{A(Li)}>1.5$, misleadingly makes the fraction of Li-rich giants to appear artificially higher: a $32\%$ of stars that have engulfed a planet have $\mathrm{A(Li)}\gtrsim 1.5$, which corresponds from $3.4\%$ to $9.6\%$ of the total population.

Low mass stars after accretion produce giants with abundances lower than $\mathrm{A(Li)}=1.5$ and would not be considered Li-rich by the criteria most commonly used, even if they are enriched. This also applies to intermediate-mass stars evolved from the Li-dip. High mass stars, on the contrary, even when evolving in isolation can have very high abundances that would be mistaken for Li-enrichment. As the population with and without SSC accretion look so similar for these stars it would be very difficult to identify stars that have gone though the engulfment process in this mass domain. Then, the currently used definition of Li-rich giant would not consider some giants that have an enhanced abundance and would mistakenly include isolated giants. This highlights the importance of studying samples with well determined evolutionary phases and masses.

Regarding the planet, the parameters that affect the most the final Li abundance of the star are the metal content and the mass, while the orbital periods almost do not change the $\mathrm{A(Li)}$. Changing the mass of the star while keeping the initial Li content constant produces changes of $0.5$ dex in Li abundance at most, but, when considering the more realistic physical domains of mass and corresponding Li abundance, the mass of the star has a large impact not only on the final Li abundance, but also on the detectability and correct identification of the engulfment signature. The metallicity of the stars is more important for the lower mass domain, where the high metallicity stars undergo larger depletion during the FDU.

The SSC engulfment scenario is a promising mechanism to explain giants in domains where extra-mixing is not expected to happen, as in metal rich stars and those located before the luminosity function bump. For these giants, the only mechanism capable of increasing the Li-abundance is pollution, and the ingestion of a SSC can raise the $\mathrm{A(Li)}$ up to $2.2$, so giants on this domain showing higher amounts of Li cannot be explained by SSC engulfment alone.
Once again, this emphasizes the importance of obtaining precise evolutionary stages for Li-rich giants to confirm what kind of mechanism could be producing the enhancement, as well as knowledge of their masses in order to know their abundance in MS and thus be able to correctly recognize the signature of the engulfment.

Ideal samples to test the SSC engulfment channel would be giants originating from the Li-dip, because of their low initial Li abundances that would allow to study a clean sample of SSC engulfment cases, and stars located below the luminosity function bump, where enhanced extra-mixing could not produce any Li enhancement. Obtaining a clear signature from the engulfment process would be challenging for high mass stars, where the engulfment of a companion almost does not change the Li abundance distribution. As an added application of our models, underlying planetary populations may be tested in mass ranges that are harder to probe with radial velocities, just by using the engulfment signature seen as an unusual abundance of Li or other chemical species.

To improve our models and make an overall more realistic analysis we need to combine our knowledge of SSC accretion with tidal evolution, extra-mixing, among others. The models presented in this work so far do not include any type of mixing that may increase (or decrease) the Li abundance, but it explores an initial picture without adding even more parameters that may complicate the interpretation of the results. Also, considering a calibrated Li turn-off abundance with mass and metallicity in our simulated distributions would generate more realistic populations to compare with observations. The numerical tools developed so far will be the base of future work where we add mixing (and production of the element) to generate a more realistic situation specially after the luminosity function bump.

\acknowledgments We thank the anonymous referee for helpful suggestions to improve the presentation of our results and quality of this paper. Support for C.A-G is provided by CONICYT-PCHA Doctorado Nacional 2013-21130353. C.A-G and JC acknowledge support from the Chilean Ministry for the Economy, Development, and Tourism's Programa Iniciativa Cient\'ifica Milenio, through grant IC120009 awarded to the Millenium Institute of Astrophysics (MAS) and from PFB-06 Centro de Astronomia y Tecnologias Afines. JC acknowledges support from Proyecto FONDECYT Regular 1130373. MHP acknowledges support from NASA grant NNX15AF13G. JKC was supported by an appointment to the NASA Postdoctoral Program at the Goddard Space Flight Center, administered by Universities Space Research Association under contract with NASA.

\bibliography{Lirichgiants_CAG}{}
\bibliographystyle{aasjournal}

\end{document}